\documentclass[12pt]{article}
\pdfoutput=1

\usepackage[square,comma,numbers,sort&compress]{natbib}
\usepackage[left=20mm,right=20mm,top=20mm,bottom=20mm]{geometry}
\usepackage{mediabb}
\usepackage{graphics}
\usepackage{epsf}
\usepackage{color}
\usepackage{amsmath}
\usepackage{amssymb}
\usepackage{latexsym}
\usepackage{slashed}
\usepackage{float}
\usepackage{cases}
\usepackage{multirow}
\usepackage{bm}
\usepackage{braket}

\newcommand{\A}{\mathcal{A}}

\newcommand{\al}[1]{\begin{align}#1\end{align}}

\newcommand{\bp}{\begin{pmatrix}}
\newcommand{\ep}{\end{pmatrix}}
\newcommand{\bb}{\begin{bmatrix}}
\newcommand{\eb}{\end{bmatrix}}

\newcommand{\del}{\partial}

\newcommand{\Ytil}{\widehat{\widetilde{Y}}}
\newcommand{\Ptil}{\widehat{\widetilde{P}}}
\newcommand{\Utwo}{\widehat{U}_{Z_2}}
\newcommand{\Uthree}{\widehat{U}_{Z_3}}
\newcommand{\Ufour}{\widehat{U}_{Z_4}}
\newcommand{\Usix}{\widehat{U}_{Z_6}}

\newcommand{\beq}{\begin{equation}}
\newcommand{\eeq}{\end{equation}}
\newcommand{\bea}{\begin{eqnarray}}
\newcommand{\eea}{\end{eqnarray}}

\newcommand{\fulltoday}{\number\day\space \ifcase\month\or
    January\or February\or March\or April\or May\or June\or
    July\or August\or September\or October\or November\or December\fi
    \space\number\year}

 \makeatletter
    \renewcommand{\theequation}{%
    \thesection.\arabic{equation}}
    \@addtoreset{equation}{section}
  \makeatother

\begin{document}
\allowdisplaybreaks[2]
\begin{titlepage}
\renewcommand\thefootnote{\alph{footnote}}
		\mbox{}\hfill EHOU-14-003\\
		\mbox{}\hfill HRI-P-14-02-001\\
		\mbox{}\hfill KOBE-TH-14-02\\
		\mbox{}\hfill KUNS-2485\\
		\mbox{}\hfill OU-HET-827\\
		\mbox{}\hfill RECAPP-HRI-2014-004\\
\vspace{4mm}
\begin{center}
{\fontsize{22pt}{0pt}\selectfont \bf {
{
Operator analysis of physical states on magnetized $T^2 / Z_{N}$ orbifolds}}} \\
\vspace{8mm}
	{\fontsize{14pt}{0pt}\selectfont \bf
	Tomo-hiro Abe,\,\footnote{
		E-mail: \tt t-abe@scphys.kyoto-u.ac.jp
		}}
	{}
	{\fontsize{14pt}{0pt}\selectfont \bf
	Yukihiro Fujimoto,\,\footnote{
		E-mail: \tt fujimoto@het.phys.sci.osaka-u.ac.jp
		}}
	{}
	{\fontsize{14pt}{0pt}\selectfont \bf
	Tatsuo Kobayashi,\,\footnote{
		E-mail: \tt kobayashi@particle.sci.hokudai.ac.jp
		}}
	\\[3pt]
	{\fontsize{14pt}{0pt}\selectfont \bf
	Takashi Miura,\,\footnote{
		{E-mail: \tt takashi.miura@people.kobe-u.ac.jp (miura\_takashi@jp.fujitsu.com)}
		}
				\footnote{Moved to \it Software Systems Laboratories, FUJITSU LABORATORIES LTD. 1-1, \\
				\textcolor{white}{spaces spaces s}Kamikodanaka 4-chome, Nakahara-ku Kawasaki, 211-8588 Japan}}
	{}
	{\fontsize{14pt}{0pt}\selectfont \bf
	Kenji Nishiwaki,\,\footnote{
		E-mail: \tt nishiwaki@hri.res.in {(nishiken@kias.re.kr)}
		}
		\footnote{{Moved to \it School of Physics, Korea Institute for Advanced Study, \\
				\textcolor{white}{spaces spaces s}85 Hoegiro, Dongdaemun-gu, Seoul, 130-722 Republic of Korea}}
		}
	{}
	{\fontsize{14pt}{0pt}\selectfont \bf
	and
	Makoto Sakamoto\,\,\footnote{
		E-mail: \tt dragon@kobe-u.ac.jp} } \\
\vspace{4mm}
	{\fontsize{13pt}{0pt}\selectfont
		${}^{\mathrm{a}}$\it Department of Physics, Kyoto University, Kyoto 606-8502, Japan \smallskip\\[3pt]
		${}^{\mathrm{b}}$\it Department of Physics, Osaka University, Toyonaka 560-0043, Japan \smallskip\\[3pt]
		${}^{\mathrm{c}}$\it Department of Physics, Hokkaido University, Sapporo 060-0810, Japan 
\smallskip\\[3pt]
		${}^{\mathrm{d,g}}$\it {Department of Physics, Kobe University, Kobe 657-8501, Japan 
\smallskip\\[3pt]
	 	${}^{\mathrm{f}}$\it Regional Centre for Accelerator-based Particle Physics,\\[2pt]
		\it Harish-Chandra Research Institute, Allahabad 211 019, India
\smallskip\\[3pt]
	}}
\vspace{4mm}
{\normalsize \fulltoday}
\vspace{10mm}
\end{center}
\begin{abstract}
{\fontsize{12pt}{16pt}\selectfont{
We discuss an effective way for analyzing the system on the magnetized twisted orbifolds in operator formalism, especially {in the complicated cases} $T^2/Z_3$, $T^2/Z_4$ and $T^2/Z_6$.
We can obtain the exact and analytical results which can be applicable for any larger values of the quantized magnetic flux {$M$, and show that the} (non-diagonalized) kinetic terms are generated via {our} formalism and {the number of the surviving physical states are calculable} in a rigorous manner by simply following usual procedures in linear algebra in any case.
{Our approach} is very powerful when we try to examine properties of the physical states on (complicated) magnetized orbifolds $T^2/Z_3, T^2/Z_4, T^2/Z_6$ (and would be in other cases on higher-dimensional torus) and could be an essential tool for actual realistic model construction based on these geometries.
}}
\end{abstract}
\end{titlepage}
\renewcommand\thefootnote{\arabic{footnote}}
\setcounter{footnote}{0}



\section{Introduction}

Even through the mass of the Higgs boson had been measured precisely at the CERN Large Hadron Collider, some other related topics are still veiled, {{\it e.g.}}, the origin of the generations, the mass hierarchy and the mixings of the Standard Model (SM) fermionic particles.
Extra dimensions can call us fascinating directions, especially when we try to solve the above problems.

Plenty of models {have} been proposed to date, and in this paper, we focus on torus compactification with magnetic flux~\cite{Bachas:1995ik,Blumenhagen:2000wh,Angelantonj:2000hi,Blumenhagen:2000ea,Cremades:2004wa}.\footnote{See for string magnetized D-brane models~\cite{Blumenhagen:2005mu,Blumenhagen:2006ci} and references therein.}
This possibility holds lots of captivating aspects.
Chiral theory can be realized as a four-dimensional low energy effective theory on the background.
Zero-mode equations are analytically solvable, apart from further nontrivial background, {{\it e.g.}}, Calabi-Yau manifolds, and interestingly, their profiles are split and quasi-localized.
The former point naturally leads to the nature with three generations, and the latter aspect would promise a natural explanation for the drastic hierarchies in the masses and the mixing patterns in the fermionic sector of the SM.\footnote{
Another possible way to tackle these problems is to introduce point interactions (zero-thickness branes) in the bulk space of a five-dimensional theory on $S^1$ and consider various boundary conditions of fields on them~\cite{Fujimoto:2012wv,Fujimoto:2013ki,Fujimoto:2014fka}.
}
Along this direction, several studies have been done to pursue (further realistic) models and to search for the phenomenological aspects on, namely, Yukawa couplings~\cite{Cremades:2004wa},\footnote{
Within the framework of superstring theory, magnetized D-brane models are T-dual of intersecting D-brane models~\cite{Blumenhagen:2005mu,Blumenhagen:2006ci}. Yukawa couplings are also computed in intersecting D-brane models~\cite{Cremades:2003qj,Cvetic:2003ch,Abel:2003vv,Honecker:2012jd}. {See also Refs.~\cite{Abel:2003yx,Abel:2005qn,Abel:2006yk,Pesando:2012cx,Pesando:2014owa}.}}
realization of quark/lepton masses and their mixing angles~{\cite{Abe:2012fj,Abe:2014vza}}, higher order couplings~\cite{Abe:2009dr}, flavor symmetries~{\cite{Abe:2009vi,Abe:2009uz,Abe:2010ii,BerasaluceGonzalez:2012vb,Honecker:2013hda,Marchesano:2013ega,Abe:2014nla}},\footnote{
A similar flavor symmetry can be obtained in heterotic string theory on an orbifold~\cite{Kobayashi:2006wq}
(see also~\cite{Kobayashi:2004ya,Ko:2007dz}).
}
massive modes~\cite{Hamada:2012wj}, and {others~\cite{Sakamoto:2003rh,Antoniadis:2004pp,Antoniadis:2009bg,Choi:2009pv,Kobayashi:2010an,DiVecchia:2011mf,Abe:2012ya,DeAngelis:2012jc,Abe:2013bba,Ferrer:1995vd,Ferrer:1994pw,DiVecchia:2007dh}.}

Another important manipulation in higher-dimensional model building is orbifolding.
By adding discrete symmetries on original backgrounds via this mechanism, we can realize supersymmetry breaking~\cite{Dixon:1985jw,Dixon:1986jc}, removing exotic particles and breaking down gauge symmetries~\cite{Kawamura:1999nj,Kawamura:2000ir,Kawamura:2000ev}.
On {two-dimensional torus} $T^2$ (without magnetic flux), not only the simplest $Z_2$ case, also more complicated twisted orbifolds $T^2/Z_N$ for $N=3,4,6$ can be constructed~\cite{Dixon:1986jc} and their geometrical aspects are discussed~\cite{Katsuki:1989bf,Kobayashi:1991rp,Choi:2006qh} within the context of string theory.
In a higher-dimensional field theory, detailed studies of $SU(N)$ and $SO(N)$ gauge theory have been carried out~\cite{Kawamura:2008mz,Kawamura:2009sa,Kawamura:2007cm,Kawamura:2009gr,Goto:2013jma}.
Furthermore on $T^6$, which has much amount of {degrees} of freedom compared with $T^2$, other complex patterns are possible like $T^6/Z_7, T^6/Z_8, T^6/Z_{12}$ and so on.

Here, we can also consider to combine the two ideas.
On orbifolded background geometries, all the states are classified under the eigenvalue (parity) of orbifold, and some zero-mode particles are projected out.
Also, mode functions are deformed from the original ones, which could be helpful when we try to realize more realistic SM flavor structure.
The simplest case of twisted orbifold with magnetic flux $T^2/Z_2$ was studied in~\cite{Abe:2008fi,Abe:2008sx},\footnote{
See for heterotic models on magnetized orbifolds~\cite{Nibbelink:2012de} and also for shifted $T^2/Z_N$ orbifold models with magnetic flux~\cite{Fujimoto:2013xha}.
}
and other cases on higher-dimensional torus (with magnetic flux) $T^6/Z_2$ and $T^6/(Z_2 \times Z'_2)$ were also already analyzed~\cite{Abe:2008fi,Abe:2008sx}.
More nontrivial twisted orbifolds on $T^2$, namely, $T^2/Z_3, T^2/Z_4, T^2/Z_6$ with magnetic flux was recently scrutinized in Ref.~\cite{Abe:2013bca}.
On these magnetized orbifolds, nontrivial (discrete) Scherk-Schwarz phases~\cite{Scherk:1978ta,Scherk:1979zr} and (discrete) Wilson line phases~\cite{Kobayashi:1991rp,Ibanez:1986tp,Kobayashi:1990mi} play an important role.\footnote{
Also in intersecting D-brane models, Scherk-Schwarz phases were discussed in~\cite{Angelantonj:2005hs} and discrete Wilson line phases were studied in~\cite{Blumenhagen:2005tn} (see also~\cite{Angelantonj:2009yj,Forste:2010gw}).
}
Especially in $T^2/Z_3$ and $T^2/Z_6$, nonzero values of Scherk-Schwarz phases and/or Wilson line phases are mandatory for defining these magnetized orbifolds consistently~\cite{Abe:2013bca}.

On the $T^2/Z_3, T^2/Z_4, T^2/Z_6$ magnetized orbifolds, constructing mode functions itself is still {formally possible} following the usual prescription.
{However,} analyzing the {number} of surviving physical states and writing down four-dimensional (4D) effective Lagrangians with suitable field normalizations are highly nontrivial because the forms of mode functions on magnetized $T^2/Z_N$ are very entangled, where we should consider (weighted) linear combinations of complicated original functions on magnetized $T^2$ which include theta functions for all the mass-degenerated states.
According to our previous analysis based on numerical computations in~\cite{Abe:2013bca},
the number of degenerated states we should consider tends to be quite large for realizing three generations, especially in the case of $T^2/Z_6$, it culminates in $24$.
Only focusing on efficiency, exact analytical evaluation of the above issues is desirable.

{In this paper,} we declare that operator formalism {gives} us a remedy {for analytic computations of important physical quantities}.
The way discussed {in operator formalism} is very powerful when we try to examine properties of the physical states on (complicated) magnetized orbifolds $T^2/Z_3, T^2/Z_4, T^2/Z_6$ and could be an essential tool for actual (realistic) model construction based on these geometries.
{
We would like to note that we can derive all the results on the number of physical modes in exact and analytic ways based on operator formalism, while it was evaluated in Ref.~\cite{Abe:2013bca} relying on (huge) numerical calculations.
}

{
This paper is constructed as follows.
{In section~\ref{section:wavefunction-analysys}, we review the previous wavefunction analysis on magnetized $T^2/Z_{N}$ orbifolds. In section~\ref{section:T2}, with keeping in mind the difficulties of the wavefunction analysis, we show how to describe the system on $T^2$ with magnetic flux in operator formalism.
Based on the knowledge, in section~\ref{section:T2/ZN}, we analyze the system on magnetized orbifolds $T^2/Z_2$ and complicated $T^2/Z_3, T^2/Z_4, T^2/Z_6$ with operator formalism step by step and lead the results in the exact way. After that, {in section~\ref{section:analysis},} we confirm the correspondence between analysis with actual forms of wavefunctions and that in operator formalism. Section~\ref{section:summary} is devoted to the conclusions and discussions.
{In Appendix~\ref{appendix:gauge}, we have a discussion on physical degrees of freedom in the quantum mechanical system in operator formalism through large gauge transformations.}
In Appendix~\ref{appendix:derivation}, derivations of the formulas which we use in section~\ref{section:T2/ZN} is supplied.}
}

{
\section{{Brief review} of the wavefunction analysis}
\label{section:wavefunction-analysys}
In this section, we review the wavefunction analysis of the magnetized $T^2 /Z_{N}$ orbifold {models} ~{\cite{Abe:2013bca}}.

In the previous wavefunction analysis, $Z_{N}$ orbifolds ($N=2,3,4,6$) were introduced {for} the torus $T^2$ with a homogeneous magnetic flux and the structure of the generations, which is a key ingredients for the generation {problem}, was evaluated. However, the analysis {totally relied} on the numerical calculations and has limitations to derive the analytic results. Moreover, there are several unclear points, {{\it e.g.,}} the meaning of the degeneracy index $j$ of the wave functions $f^{(j+\alpha_{1},\alpha_{\tau})}_{T^2,\psi_{+},0}(z;a_{w})$, $g^{(j+\alpha_{1},\alpha_{\tau})}_{T^2,\psi_{+},0}(z;a_{w})$ {(see Eq.~(\ref{fermion_mode_function}))}.  {The index $j$ is expected to be an eigenvalue of an operator though no one finds such an operator so far. }

As we will see in this section, the difficulties to derive the exact results appear from the $Z_{N}$-rotated zero-mode functions. On the magnetized $T^2 /Z_{N}$ orbifold, the physical states consist of the proper linear combinations of the $Z_{N}$-rotated zero modes. The $Z_{N}$-rotated zero modes also can be expanded by the original zero modes {on $T^2$} since both of them satisfy the same zero-mode equation and the boundary conditions. {Then, we need to compute the expansion coefficients of the $Z_{N}$-rotated states to find the number of physical states on $T^2/Z_{N}$.} However, the derivation of the {expansion coefficients} with the general magnetic flux is {very} difficult {in an analytic way} because the zero-mode functions contain {theta functions}.

We will see in the section~\ref{section:T2/ZN} that, in the operator analysis, we can overcome {these} difficulties and can obtain {exact} analytic results for the number of physical states and the {expansion coefficients}. The correspondence between the wavefunction analysis and the operator analysis can be found in {section~\ref{section:analysis}}.
}

\subsection{Fermion zero-mode wavefunctions on $T^2$ and $T^2/Z_N$ with magnetic flux}

We consider the 6D action on $M^4 \times T^2$ with magnetic flux~\cite{Cremades:2004wa,Hashimoto:1997gm}
\al{
\int_{M^4} d^4 x \int_{T^2} dz d\bar{z} \left\{ i \bar{\Psi}_{+} \Gamma^M D_M \Psi_{+} \right\},
\label{6D_fermion_action}
}
where the capital roman indices $M,N$ run over $\mu\,(=0,1,2,3)$, $z$ and $\bar{z}$.
{
The complex coordinate $z$ ($\bar{z}$), which is very useful for evaluating the actual forms of wavefunctions, is defined as $z = y_1 + i y_2$ ($\bar{z} = y_1 - i y_2$), where $y_{1}$ and $y_{2}$ are {two Cartesian coordinates. Here, we take the bases of the torus as $\mathbf{u}_1=(1,0)^{\text{T}}$, $\mathbf{u}_2 =(\text{Re}\,\tau ,\text{Im}\,\tau)^{\text{T}}$} with the modulus parameter $\tau$\,$(\tau \in \mathbb{C},\ \text{Im}\tau > 0)$ for convenience.}


$\Psi_+$ is a 6D Weyl fermion with 6D positive chirality and the covariant derivative $D_M \,(:= \partial_M -iq A_M)$ represents the gauge interaction with a $U(1)$ gauge field $A_M$ with the background configurations $A_{z}^{(b)}$ and $A_{\bar{z}}^{(b)}${, where $q$ is a $U(1)$ charge.}
On $T^2$, the complex coordinate $z$ is identified as $z \sim z+1 \sim z+\tau$, the counterpart of which in operator formalism is found in Eq.~(\ref{torus_definition}).
When we consider the case with a 6D Weyl fermion with 6D negative chirality $\Psi_{-}$, the resultant 4D chirality is simply flipped.
We use the notations for representations of 6D Clifford algebra and complex coordinates adopted in Ref.~\cite{Abe:2013bca}.

The vector potential $A^{(b)}$ describing the magnetic flux {$b=\int_{T^2} F$ through the field strength $F = \frac{ib}{2\mathrm{Im}\tau} dz \wedge d\bar{z}$} can be written as
\begin{align}
A^{(b)}(z)
&={b\over 2\mathrm{Im}\tau}\mathrm{Im}[(\bar{z}+\bar{a}_w)dz] \notag \\
&={b\over 4i\mathrm{Im}\tau}(\bar{z}+\bar{a}_w)dz-{b\over 4i\mathrm{Im}\tau}(z+a_w)d\bar{z}  \notag \\[2mm]
&=:
{A_{z}^{(b)}(z)\hspace{0.5mm}dz+A_{\bar{z}}^{(b)}(z)\hspace{0.5mm}d\bar{z}
  \mathstrut} ,
\label{v_pA}
\end{align}
where $a_w$ is a complex Wilson line phase.
{From Eq.~(\ref{v_pA}), we obtain} 
\begin{align}
A^{(b)}(z+1)&=A^{(b)}(z)+{b\over 2\mathrm{Im}\tau}\mathrm{Im}dz 
=: A^{(b)}(z)+d\chi_1 (z+a_w),~~\notag  \\
A^{(b)}(z+\tau )&=A^{(b)}(z)+{b\over 2\mathrm{Im}\tau}\mathrm{Im}(\bar{\tau}dz) 
=: A^{(b)}(z)+d\chi_{\tau} (z+a_w),
\label{v_pAtra}
\end{align}
where $\chi_1(z+a_w)$ and $\chi_{\tau}(z+a_w)$ are given by\footnote{
Note that we can freely add constants in the definition of $\chi_1$ and $\chi_{\tau}$ without changing the relation (\ref{v_pAtra}).
Here, 
we chose such constants, as in Eq.~(\ref{def of chi}), for later convenience.}
\begin{align}
\chi_1(z+a_w)={b\over 2\mathrm{Im}\tau}\mathrm{Im}(z+a_w),~~
\chi_{\tau}(z+a_w)={b\over 2\mathrm{Im}\tau}\mathrm{Im}[\bar{\tau}(z+a_w)].
\label{def of chi}
\end{align}
Here, the Lagrangian density {in} Eq.~(\ref{6D_fermion_action}) should be single-valued under the identification $z \sim z+1 \sim z+\tau$, and thereby {the} field $\Psi_+(x,z)$ should satisfy the {pseudo-periodic} boundary conditions
\begin{align}
\Psi_+(x,z+1) =U_1(z)\Psi_+(x,z),~~
\Psi_+(x,z+\tau) =U_{\tau}(z)\Psi_+(x,z), 
\label{Psi_M^4T^2}
\end{align}
with
\begin{align}
U_1(z) := e^{iq\chi_1(z+a_w) +2\pi i \alpha_1}, ~~~U_{\tau}(z) := e^{iq\chi_{\tau}(z+a_w) +2\pi i \alpha_{\tau}},
\label{transformations for lattice shifts}
\end{align}
where $\alpha_1$ and $\alpha_{\tau}$ are allowed to be any real {numbers}, and {correspond to} Scherk-Schwarz phases.
{The consistency of the contractible loops,} {{\it e.g.}}, $z\to z+1\to z+1+\tau\to z+\tau\to z$, requires the magnetic flux quantization condition,
\begin{align}
{qb \over 2\pi} =: M\in \mathbb{Z}.
\end{align}
Then, $U_1(z)$ and $U_{\tau}(z)$ satisfy 
\begin{align}
U_1(z+\tau)U_{\tau}(z) =U_{\tau}(z+1)U_1(z).
\end{align}

It should be emphasized that all of the Wilson line {phases} and the
Scherk-Schwarz phases can be arbitrary {on $T^2$}, but are not physically
independent because the Wilson line {phases} can be absorbed into the
Scherk-Schwarz phases by a redefinition of fields and 
vice versa.
This fact implies that we can take, for instance, the basis of vanishing Wilson line phases, without any loss of generality.
It is then interesting to point out 
{that }allowed Scherk-Schwarz phases
are severely restricted for $T^2/Z_N$ {orbifold models,} as we {will see in section~\ref{section:T2/ZN}}, 
while there is no restriction on the Scherk-Schwarz phases for $T^2$ models. 

It is known that the zero-mode {states} of $\Psi_+$ {become} chiral and multiple due to the effect of the magnetic flux
\al{
\Psi_{+,0}(x,z) =
\begin{cases}
	\displaystyle
	\sum_{j=0}^{|M|-1}
	\begin{pmatrix} \psi_{R,0}^{(j)}(x) \\ 0 \end{pmatrix}
	f_{T^2,{\Psi_+},0}^{(j+\alpha_1,\alpha_\tau)} (z;a_w) &
	\text{for } M>0, \\
	\displaystyle
	\sum_{j=0}^{|M|-1}
	\begin{pmatrix} 0 \\ \psi_{L,0}^{(j)}(x) \end{pmatrix}
	{g}_{T^2,{\Psi_+},0}^{(j+\alpha_1,\alpha_\tau)} (z;a_w) &
	\text{for } M<0,
\end{cases}
\label{fermion_KK_expansion}
}
where $j=0,1,\cdots, |M|-1$ {is just {an} index for {$|M|$-}degenerated states {and} $\psi_{R,0}^{(j)}(x)$ or $\psi_{L,0}^{(j)}(x)$ are 4D chiral zero modes.
The mode functions {{$f_{T^2,{\Psi_+},0}^{(j+\alpha_1,\alpha_\tau)} (z;a_w)$ and ${g}_{T^2,{\Psi_+},0}^{(j+\alpha_1,\alpha_\tau)} (z;a_w)$}} {are the solutions of the zero-mode {equations}
	\begin{align}
	\left(\partial_{\bar{z}}+\frac{\pi M}{2{\rm Im}\,\tau}(z+a_{w})\right)f_{T^2,{\Psi_+},0}^{(j+\alpha_1,\alpha_\tau)} (z;a_w)=0,\notag\\
	\left(\partial_{z}-\frac{\pi M}{2{\rm Im}\,\tau}(\bar{z}+\bar{a}_{w})\right){g}_{T^2,{\Psi_+},0}^{(j+\alpha_1,\alpha_\tau)} (z;a_w)=0, \label{zero-mode-equation}
	\end{align}
and} take the forms
\al{
f_{T^2,{\Psi_+},0}^{(j+\alpha_1,\alpha_\tau)} (z;a_w) &=
\mathcal{N} e^{i\pi M (z+a_w){\mathrm{Im}(z+a_w)\over \mathrm{Im}\tau}}\cdot \vartheta \left[
\begin{array}{c}
{j+\alpha_1 \over M} \\ -\alpha_{\tau}
\end{array}
\right] (M(z+a_w),M\tau), \notag \\
{g}_{T^2,{\Psi_+},0}^{(j+\alpha_1,\alpha_\tau)} (z;a_w) &=
\mathcal{N} e^{i\pi M (\bar{z}+\bar{a}_w){\mathrm{Im}(\bar{z}+\bar{a}_w)\over \mathrm{Im}\bar{\tau}}}\cdot \vartheta \left[
\begin{array}{c}
{j+\alpha_1 \over M} \\ -\alpha_{\tau}
\end{array}
\right] (M(\bar{z}+\bar{a}_w),M\bar{\tau}),
\label{fermion_mode_function}
}
with the normalization factor $\mathcal{N}$.
For $a_w=0$ and $(\alpha_1,\alpha_{\tau})=(0,0)$,
$\psi_{\pm,0}^{(j+\alpha_1,\alpha_{\tau})}(z;a_w)$ are reduced to the
results obtained 
in Ref.~\cite{Cremades:2004wa}.
Here, $\mathcal{N}$ may be fixed by the orthonormality condition
{\begin{align}
\int_{T^2}dzd\bar{z}~\left(f_{T^2,\Psi_+,0}^{(j+\alpha_1,\alpha_{\tau})} (z;a_w) \right)^\ast
f_{T^2,\Psi_+,0}^{(k+\alpha_1,\alpha_{\tau})} (z;a_w) &= \delta_{jk}, \notag \\
\int_{T^2}dzd\bar{z}~\left(g_{T^2,\Psi_+,0}^{(j+\alpha_1,\alpha_{\tau})} (z;a_w) \right)^\ast
g_{T^2,\Psi_+,0}^{(k+\alpha_1,\alpha_{\tau})} (z;a_w) &= \delta_{jk}.
\label{T2_orthonormal_relation}
\end{align}}

{It should be stressed that although $j$ ($=0,1,2,\cdots,|M|-1$) in Eq.~(\ref{fermion_mode_function}) is the index that distinguishes the $|M|$ degenerate zero-modes {and is expected an eigenvalue of some hermitian operator, the form of such operator is unclear} in the wavefunction approach. {We will later clarify the operator, which is crucial to evaluate the exact analytic results, in operator formalism.}} 

{The }$\vartheta$ function is defined by 
\begin{align}
&\vartheta \left[
\begin{array}{c}
a\\ b
\end{array}
\right] (c\nu,c\tau) 
=\sum_{l=-\infty}^{\infty}e^{i\pi (a+l)^2c\tau}e^{2\pi i(a+l)(c\nu +b)} ,
\end{align}
with the properties
\begin{align}
&\vartheta \left[
\begin{array}{c}
a\\ b
\end{array}
\right] (c(\nu +n),c\tau)
=e^{2\pi i acn}\vartheta \left[
\begin{array}{c}
a\\ b
\end{array}
\right] (c\nu,c\tau), \notag \\
&\vartheta \left[
\begin{array}{c}
a\\ b
\end{array}
\right] (c(\nu +n\tau),c\tau)
=e^{-i\pi cn^2\tau -2\pi i n(c\nu +b)}\vartheta \left[
\begin{array}{c}
a\\ b
\end{array}
\right] (c\nu,c\tau), \notag \\
&\vartheta \left[
\begin{array}{c}
a+m\\ b+n
\end{array}
\right] (c\nu,c\tau)
=e^{2\pi i an}\vartheta \left[
\begin{array}{c}
a\\ b
\end{array}
\right] (c\nu,c\tau), 
\end{align}
where $a$ and $b$ are real numbers, $c$, $m$ and $n$ are integers, and $\nu$ and $\tau$ are complex numbers with $\mathrm{Im}\tau >0$.

Now, we explicitly write down a part of the 4D effective Lagrangian describing fermion kinetic terms.
Through Eqs.~(\ref{fermion_KK_expansion}),
(\ref{fermion_mode_function}) and (\ref{T2_orthonormal_relation}),
when $M>0$,
Eq.~(\ref{6D_fermion_action}) leads to the following {zero-mode} part
\al{
\sum_{j,k=0}^{|M|-1} \left\{ i \bar{\psi}_{R,0}^{(j)} [\delta_{jk}] \gamma^{\mu} \del_{\mu} \psi_{R,0}^{(k)} \right\}
=
\sum_{j=0}^{|M|-1} \left\{ i \bar{\psi}_{R,0}^{(j)} \gamma^{\mu} \del_{\mu} \psi_{R,0}^{(j)} \right\},
}
where we obtain a $|M|$-generation chiral theory.
When we consider $M<0$, the {chirality} of the fermions {turns} out to be left-handed.

After imposing $Z_N$-orbifolding on $T^2$ with $\omega = e^{2\pi i/N}$ in which the modulus parameter $\tau$ was identified with $\omega$ for $T^2/Z_3$, $T^2/Z_4$ and $T^2/Z_6$, it had been found that the allowed Scherk-Schwarz phases ($\alpha_{1},\alpha_{\tau}$) are restricted to the specific values in the basis of $a_{w}=0$~{\cite{Abe:2013bca}}. We should mention that we can always move to the basis in which the Wilson line phase {$a_{w}$} vanishes without any loss of generality through a large gauge transformation. Thus, in the following, we concentrate on the situation
\al{
a_{w}=0.
}
On the orbifolds, the $Z_N$-orbifolded mode {functions,
\begin{align}
f_{T^2/Z_N,{\Psi_+},{0}}^{(j+\alpha_1,\alpha_{\tau})}(z;0)_{\eta}
&=\mathcal{N}^{(j)}_{R,\eta}\sum_{x =0}^{N-1}\bar{\eta}^{x} f_{T^2,{\Psi_+},{0}}^{(j+\alpha_1,\alpha_{\tau})}(\omega^{x}z;0), 
\notag \\
{g}_{T^2/Z_N,{\Psi_+},{0}}^{(j+\alpha_1,\alpha_{\tau})}(z;0)_{\omega\eta}
&=\mathcal{N}^{(j)}_{L,\omega \eta}\sum_{x =0}^{N-1}(\bar{\omega}\bar{\eta})^{x} {g}_{T^2,{\Psi_+},{0}}^{(j+\alpha_1,\alpha_{\tau})}(\omega^{x}z;0), 
\label{ZNorbifoldedmodefunctions}
\end{align}
with} the normalizing factors $\mathcal{N}^{(j)}_{R,\eta} = \mathcal{N}^{(j)}_{L,\omega \eta} = {1/N}$,\footnote{
This factor depends on a choice of the range of the integration $\int dz d\bar{z}$.
${1/N}$ corresponds to the case $\int_{T^2} dz d\bar{z}$ (after the orbifolding).
} fulfill the following conditions{:
\begin{align}
f^{(j+\alpha_1,\alpha_{\tau})}_{T^2/Z_N,{\Psi_+},{0}}(\omega z; 0)_{\eta} &= \eta f^{(j+\alpha_1,\alpha_{\tau})}_{T^2/Z_N,{\Psi_+},{0}}(z; 0)_{\eta}, \notag \\
g^{(j+\alpha_1,\alpha_{\tau})}_{T^2/Z_N,{\Psi_+},{0}}(\omega z; 0)_{\omega\eta} &= \omega \eta \,g^{(j+\alpha_1,\alpha_{\tau})}_{T^2/Z_N,{\Psi_+},{0}}(z; 0)_{\omega\eta},
\end{align}
where} $\eta$ is {one of} the possible eigenvalues {$\eta \in \{1,\omega,\omega^2,\cdots, \omega^{N-1} \}$}. {Here, the {subscript} $\eta$ of the function $f^{(j+\alpha_1,\alpha_{\tau})}_{T^2/Z_N,{\Psi_+},{0}}(\omega z; 0)_{\eta}$ denotes the eigenvalue of $f^{(j+\alpha_1,\alpha_{\tau})}_{T^2/Z_N,{\Psi_+},{0}}(\omega z; 0)_{\eta}$ under the $Z_{N}$ rotation: $z \rightarrow \omega z$.} 

{We} note that the $Z_{N}$-rotated mode function {$f^{(j+\alpha_1,\alpha_{\tau})}_{T^2,{\Psi_+},{0}}(\omega^x z;0)_{\eta} $  $\Bigl(g^{(j+\alpha_1,\alpha_{\tau})}_{T^2,{\Psi_+},{0}}(\omega^x z; 0)_{\omega\eta}\Bigr)$} in Eq.~(\ref{ZNorbifoldedmodefunctions}) can be expanded by the original mode functions {as
\al{
f^{(j+\alpha_1,\alpha_{\tau})}_{T^2,{\Psi_+},{0}}(\omega^x z; 0)=\sum_{k=0}^{|M|-1}C^{(\omega^{x})}_{jk}f^{(k+\alpha_1,\alpha_{\tau})}_{T^2, {\Psi_+},{0}}(z; 0) ,
}
with} the {expansion coefficients
\al{
C_{jk}^{(\omega^x)}=\int dzd\bar{z}\ f^{(j+\alpha_1,\alpha_{\tau})}_{T^2,{\Psi_+},{0}}(\omega^x z; 0)\Bigl(f^{(k+\alpha_1,\alpha_{\tau})}_{T^2, {\Psi_+},{0}}(z; 0)\Bigr)^{\ast},\label{expandcoeffient}
}
since} they satisfy the same zero-mode equation (\ref{zero-mode-equation}) and the boundary conditions (\ref{Psi_M^4T^2}). {As in the $T^2$ case of Eq.~(\ref{fermion_KK_expansion})}, the zero modes on $T^2/Z_N$ are represented as follows:
\al{
\Psi_{+,0}(x,z) =
\begin{cases}
	\displaystyle
	\sum_{j=0}^{|M|-1}
	\begin{pmatrix} \psi_{R,0}^{(j)}(x) \\ 0 \end{pmatrix}
	f_{T^2/Z_N,{\Psi_+},0}^{(j+\alpha_1,\alpha_\tau)} (z;0)_{\eta} &
	\text{for } M>0, \\
	\displaystyle
	\sum_{j=0}^{|M|-1}
	\begin{pmatrix} 0 \\ \psi_{L,0}^{(j)}(x) \end{pmatrix}
	{g}_{T^2/Z_N,{\Psi_+},0}^{(j+\alpha_1,\alpha_\tau)} (z;0)_{\omega\eta} &
	\text{for } M<0.
\end{cases}
}
In the $T^2/Z_N$ cases with {$\eta$}, the fermion kinetic terms could be evaluated like the previous $T^2$ case {as
\al{
\sum_{j,k=0}^{|M|-1} \left\{ i \bar{\psi}_{R,0}^{(j)} [\mathcal{K}_{jk}^{(Z_{N};\eta)}] \gamma^{\mu} \del_{\mu} \psi_{R,0}^{(k)} \right\}{,}
}
with} {the kinetic matrix $\mathcal{K}_{jk}^{(Z_{N};\eta)}$
\al{
\mathcal{K}_{jk}^{(Z_{N};\eta)} &= 
\int_{T^2}dzd\bar{z}~\left(f_{T^2/Z_N,{\Psi_+},0}^{(j+\alpha_1,\alpha_{\tau})} (z)_{\eta} \right)^\ast
f_{T^2/Z_N,{\Psi_+},0}^{(k+\alpha_1,\alpha_{\tau})} (z)_{\eta}\nonumber\\
&=\frac{1}{N^2}\sum^{N-1}_{x=0}\sum^{N-1}_{y=0}\, \eta^{x}\bar{\eta}^{y} \sum^{|M|-1}_{m=0}\left(C^{(\omega^x)}_{jm}\right)^{\ast}C_{km}^{(\omega^y)}.
\label{kinetic_matrix}
}
The matrix $\mathcal{K}_{jk}^{(Z_{N};\eta)}$} is generally non-diagonal $|M|$-by-$|M|$ matrix because the $T^2/Z_N$ mode functions are constructed as linear combinations of {$Z_{N}$-rotated zero-mode functions on $T^2$ {(see Eq.~(\ref{ZNorbifoldedmodefunctions}))}}.
{An important point is that the rank of the matrix {$\mathcal{K}_{jk}^{(Z_{N};\eta)}$} indicates the number of physical states on $T^2 / Z_{N}$ orbifold:
	\al{
	\text{The number of physical states}= {\rm Rank}\left[{\mathcal{K}_{jk}^{(Z_{N};\eta)}}\right]. \label{RankK}
	}
To estimate the matrix {$\mathcal{K}_{jk}^{(Z_{N};\eta)}$} with general $|M|$, we need to evaluate the {expansion coefficients} (\ref{expandcoeffient}) thereby the integral Eq.~(\ref{kinetic_matrix}) can be {executed. However,} it is enormously difficult to do it {for large $|M|$} in the wavefunction analysis because of the existence of the theta functions (\ref{fermion_mode_function}){, though numerical calculations of Eq.~(\ref{expandcoeffient}) could be performed for small $|M|$}.  

As we will see in section~\ref{section:T2/ZN}, in the operator analysis, we can overcome {these} difficulties and can obtain the exact analytic results for the number of physical states and the {expansion coefficients}. {The hermitian operator, which has relation to the degeneracy index $j$ in the wavefunction (\ref{fermion_mode_function}),} also becomes clear in Section~\ref{section:T2}.  }


\section{Operator formalism for $T^2$ with magnetic flux
\label{section:T2}}

Before we consider {$Z_{N}$-}orbifolds, we formulate a quantum mechanical system on $T^2$ with a $U(1)$ homogeneous magnetic flux like in {Ref.~\cite{Abe:2013bca}}.
Its energy spectrum should correspond to the mass spectrum of the {six-dimensional (6D)} system $M^4 \times T^2$, where $M^4$ means {four-dimensional (4D)} Minkowski spacetime.
Due to compactness of the system, we face to additional constraints on the system, part of which describes degeneracy of the allowed physical states.

At first, we describe the system with the wavefunction $\psi(\mathbf{y})$, where we adopt the vector notation on $T^2$ as $\mathbf{y} := (y_1, y_2)^{\mathrm{T}}$ for two Cartesian coordinates $y_1$ and $y_2$.
We consider the Hamiltonian $H$ and the corresponding {Schr\"{o}dinger} equation with energy $E$ as
\al{
H = \left( -i \nabla {-} q \mathbf{A}(\mathbf{y}) \right)^2,\quad
H \psi(\mathbf{y}) = E \psi(\mathbf{y}),
		\label{original_Hamiltonian}
}
where we use the vector notations of $\nabla := (\partial_{y_1}, \partial_{y_2})^{\mathrm{T}}$ and $\mathbf{A} := (A_{y_1}, A_{y_2})^{\mathrm{T}}${.}
The vector potential $\mathbf{A}$ providing a homogeneous magnetic flux penetrating $T^2$ can be expressed as
\al{
\mathbf{A}(\mathbf{y}) = - \frac{1}{2} \Omega {\left( \mathbf{y} + \mathbf{a} \right)},
}
{or} with showing all the indices explicitly
\al{
A_i(\mathbf{y}) = - \frac{1}{2} \sum_{j=1}^{2} \Omega_{ij} {\left( y_j +a_j \right)},\quad
(i = 1,2).
}
{Here}, we mention that only the antisymmetric part of $\Omega$ possesses {a} physical degree of freedom, while the symmetric part of {$\Omega$} depending on a choice of gauge is unphysical and can be gauged away.
In a later stage, we suitably fix the gauge to make the system simplified.

{Two-dimensional torus} $T^2$ is defined from the two-dimensional plane $\mathbb{R}^2$ by modding out the lattice shift $\Lambda$ with {two} basis vectors $\mathbf{u}_1$ and $\mathbf{u}_2$:
\al{
T^2 = \mathbb{R}^2/\Lambda,\quad
\Lambda = \Bigg\{ \sum_{a=1}^{2} n_a \mathbf{u}_a \Big| n_a \in \mathbb{Z} \Bigg\},
\label{lattice_shift}
} 
where $n_1$ and $n_2$ {show} the {numbers} of shifting along $\mathbf{u}_1$ and $\mathbf{u}_2$ {directions}, respectively.
This means that the vector coordinate $\mathbf{y}$ should obey the identification
\al{
\mathbf{y} \sim \mathbf{y} + \sum_{a=1}^{2} n_a \mathbf{u}_a.
\label{torus_definition}
}

The above condition leads to two requirements for ensuring consistency.
The first one {puts} a constraint on the form of the wavefunction after shifting $\psi(\mathbf{y} + \mathbf{u}_a)$.
In order to make the Schr\"{o}dinger equation {well-defined} on $T^2$,
the following pseudo-periodic boundary condition should be satisfied,
\al{
\psi(\mathbf{y} + \mathbf{u}_a) =
	e^{{-}i \frac{q}{2} (\mathbf{y}+\mathbf{a})^{\mathrm{T}} \Omega \mathbf{u}_a + 2\pi i \alpha_a} \psi(\mathbf{y})\quad
	\text{for } a=1,2,
	\label{constraint_on_torus_wavefunction}
}
{where} $\alpha_a$ shows the $y$-independent Scherk-Schwarz phase {for} the $\mathbf{u}_a$ direction.\footnote{
In this paper, the Scherk-Schwarz phases only represent twisted boundary conditions, not supersymmetry breaking.}
In addition, after considering contractible loops on $T^2$,
{{{\it e.g.}}, $\mathbf{y} \to \mathbf{y}+\mathbf{u}_1 \to \mathbf{y}+\mathbf{u}_1+\mathbf{u}_2 \to \mathbf{y}+\mathbf{u}_2 \to \mathbf{y}$,}
the magnetic flux should be quantized to ensure the single-valuedness of the wavefunction $\psi(\mathbf{y})$ as follows:
\al{
q \mathbf{u}_a^{\mathrm{T}} B \mathbf{u}_b = 2 \pi Q_{ab},\quad
{(a,b = 1,2)},
\label{magnetic_flux_quantization}
}
where {$B:=\frac{1}{2} (\Omega - \Omega^{\mathrm{T}})$ is the gauge-independent antisymmetric part of $\Omega$}
and $Q_{ab} = - Q_{ba} \in \mathbb{Z}$.

Now, we go {to} the operator formalism, where we introduce a momentum operator $\widehat{\mathbf{p}} := -i \nabla$ being conjugate to $\mathbf{y}$.
The operators satisfy the canonical commutation relations,
\al{
{\left[ \widehat{y}_i, \widehat{p}_j \right]} = i \delta_{i,j},\quad
(\text{the others}) = 0,\quad
(i,j = 1,2).
}
The wavefunction $\psi(\mathbf{y})$ is represented in the operator formulation as $\braket{\mathbf{y} | \psi}$ and the system is rewritten by use of $\ket{\psi}$ as
\al{
\widehat{H} = \left( \widehat{\mathbf{p}} {+} \frac{q}{2} \Omega {\left( \widehat{\mathbf{y}} + \mathbf{a} \right)} \right)^2 {=:} (\widehat{\mathbf{p}}')^2,\quad
\widehat{H} \ket{\psi} = E \ket{\psi}.
		\label{system_operator}
}
{The} constraints in Eq.~(\ref{constraint_on_torus_wavefunction}) are also interpreted as
\al{
e^{i \widehat{T}_a {-} i \frac{q}{4} \mathbf{u}_a^{\mathrm{T}} \Omega \mathbf{u}_a } \ket{\psi} = e^{2\pi i \alpha_a} \ket{\psi},\qquad
{
	\widehat{T}_a = {\mathbf{u}_a^{\mathrm{T}} \left( \widehat{\mathbf{p}} {+}  \frac{q}{2} \Omega^{\mathrm{T}} (\widehat{\mathbf{y}}+\mathbf{a}) \right)},\quad}
	(a=1,2).
		\label{constraint_on_torus_states}
}
Here, we can find the relation
\al{
\mathbf{u}_a^{\mathrm{T}} \Omega \mathbf{u}_a =
\mathbf{u}_a^{\mathrm{T}} \left( \frac{1}{2} (\Omega + \Omega^{\mathrm{T}}) \right) \mathbf{u}_a,
		\label{transformation_for_symmetricform}
}
which says that only unphysical components appear in this part.
Hence in the following part, we drop the unphysical symmetric part of $\Omega$,
\al{
\frac{1}{2} \left( \Omega + \Omega^{\mathrm{T}} \right) = 0,
\label{gauge_fixing}
}
which leads to the condition $\mathbf{u}_a^{\mathrm{T}} \Omega \mathbf{u}_a = 0$.
We can define new operators,
\al{
&\widehat{Y} = {-}\frac{\sqrt{2}}{\omega} \widehat{p}'_2,\quad
\widehat{P} = \sqrt{2} \widehat{p}'_1,\\
&\widehat{\widetilde{Y}} = \frac{1}{2 \pi M} \widehat{T}_1 {-} \frac{\alpha_1}{M},\quad
\widehat{\widetilde{P}} = -\widehat{T}_2 + 2\pi \alpha_2,
		\label{new_operators}
}
{where} {$\omega=2q B_{12}=2q\Omega_{12}$} and $M := Q_{12} \subset \mathbb{Z}$.
The transformation $\{ \widehat{y}_i, \widehat{p}_i;\ i=1,2 \} \to \{ \widehat{Y}, \widehat{P}, \widehat{\widetilde{Y}}, \widehat{\widetilde{P}} \}$ is canonical and then the operators are suitably quantized with the canonical commutation relations,
\al{
\left[ \widehat{Y}, \widehat{P} \right] = i,\quad
{\Big[} \widehat{\widetilde{Y}}, \widehat{\widetilde{P}} {\Big]} = i,\quad
(\text{the others}) = 0.
		\label{new_commutation_relations}
}

As apparent from Eqs.~(\ref{system_operator}) and (\ref{new_operators}), under the new variables, the Hamiltonian can be rephrased only with the two new variables $\widehat{P}$ and $\widehat{Y}$ as
\al{
\widehat{H} = \frac{1}{2} \widehat{P}^2 + \frac{\omega^2}{2} \widehat{Y}^2,
		\label{Hamiltonian_newoperator}
}
the form of which is the one-dimensional harmonic oscillator.
On the other hand, the remaining two ones, $\widehat{\widetilde{P}}$ and $\widehat{\widetilde{Y}}$, work as constraint conditions on the state,
\al{
e^{i \widehat{\widetilde{P}}} \ket{\psi} = \ket{\psi},\quad
{e^{2\pi i M \widehat{\widetilde{Y}}}} \ket{\psi} = \ket{\psi},
		\label{constraints_on_states}
}
where the two operators are considered to {control} degenerated states since they do not appear in the Hamiltonian~(\ref{Hamiltonian_newoperator}) {and they commute with the Hamiltonian~(\ref{Hamiltonian_newoperator})}.
In the following part, we check this statement.

After we take an eigenstate of $\widehat{\widetilde{Y}}$, which obeys the relation $\widehat{\widetilde{Y}} \ket{\widetilde{Y}} = \widetilde{Y} \ket{\widetilde{Y}}$,
the second condition in Eq.~(\ref{constraints_on_states}) is simplified,
\al{
{e^{2\pi i M \widetilde{Y}} \ket{\widetilde{Y}} =
\ket{\widetilde{Y}}}.\label{Y-1}
}
Operating ${e^{ 2\pi i M \widehat{\widetilde{Y}}}}$ on the state $e^{i a \widehat{\widetilde{P}}} \ket{\widetilde{Y}}\, (a \in \mathbb{R})$ and using the relations in Eq.~(\ref{new_commutation_relations}),
we can obtain the {relation}
\al{
e^{i a \widehat{\widetilde{P}}} \ket{\widetilde{Y}} = \ket{\widetilde{Y} - a}.
		\label{form_of_transration}
}
From Eqs.~(\ref{constraints_on_states}), {(\ref{Y-1})} and (\ref{form_of_transration}), we can reach the periodic condition,
\al{
\Ket{\widetilde{Y}} = \Ket{\widetilde{Y}-1},
		\label{periodic_condition}
}
and {also} the coordinate quantization condition,
\al{
\widetilde{Y} = \frac{j}{M}, \quad (j = 0,1,2,\cdots,|M|-1).
		\label{coordinate_quantization_condition}
}
These results imply that every energy state is $|M|$-fold degenerated and the index $j$ discriminates them {as
\al{
\widehat{H} \Ket{n, \frac{j}{M}} = E_n \Ket{n, \frac{j}{M}},\quad
E_n = |\omega| \left( n + \frac{1}{2} \right),
}
\vspace{-4mm}
\al{
\Braket{m,\frac{i}{M}|n,\frac{j}{M}} = \delta_{m,n} \delta_{i,j},
}
where} $m,n=0,1,2,\cdots$ and $i,j = 0,1,2,\cdots,|M|-1$.

{Here, {we} note three things}.
{The first one} is that there is no constraint on the Scherk-Schwarz phases $\alpha_1$ and $\alpha_2$ on the two-dimensional torus with magnetic flux. {We will see in Section~\ref{section:T2/ZN} that only the restricted values are allowed for the Scherk-Schwarz phase in the case of magnetized $T^2/Z_{N}$ orbifolds. 

{The second one} is that the index $j$ for the degeneracy states is nothing but {an} eigenvalue of the operator $\widehat{\widetilde{Y}}$. We can easily check that the index $j$ in wavefunction analysis is also the eigenvalue of  $\widehat{\widetilde{Y}}$:
	\al{
	e^{2\pi i \widehat{\widetilde{Y}}}f^{(j+\alpha_{1},\alpha_{\tau})}_{T^2,\Psi_{+},0}(z;a_{w})=e^{2\pi i \frac{j}{M}}f^{(j+\alpha_{1},\alpha_{\tau})}_{T^2,\Psi_{+},0}(z;a_{w}),\label{Ytilwavefunction}
	}
where the explicit form of the operator {$\widehat{\widetilde{Y}}$} in the complex coordinate is given by the following{,}
	\al{
	e^{2\pi i \widehat{\widetilde{Y}}}=e^{ 2\pi i \cdot \frac{1}{2\pi M}\left(-i(\partial_{z}+\partial_{\bar{z}})-\frac{\pi M}{{\rm Im}\tau}{\rm Im}\, (z+a_{w})-2\pi\alpha_{1}\right)}.
	}
On the other hand, the operator {$\widehat{\widetilde{P}}$} {acts as} the translational operator with respect to the index $j${,}
	\al{
	e^{-i\frac{1}{M}\widehat{\widetilde{P}}}f^{(j+\alpha_{1},\alpha_{\tau})}_{T^2,\Psi_{+},0}(z;a_{w})=f^{\left((j+1)+\alpha_{1},\alpha_{\tau}\right)}_{T^2,\Psi_{+},0}(z;a_{w}),\label{Ptilwavefunction}
	}
where the explicit form of the operator {$\widehat{\widetilde{P}}$} in the complex coordinate {is
	\al{
	e^{-i\frac{1}{M}\widehat{\widetilde{P}}}=e^{-i\frac{1}{M}\left(\,i(\tau\partial_{z}+\bar{\tau}\partial_{\bar{z}})-\frac{\pi M}{{\rm Im}\tau}{\rm Im}[\tau (\bar{z}+\bar{a}_{w})]+2\pi\alpha_{2}\right)}.
	}
Thus, we have succeeded to clarify {the form of the hermitian operator which has relation to} the index $j$, as announced before. 

The last one} is that $E_n$ {correspond} to eigenvalues of the Laplace operator with magnetic flux in Eq.~(\ref{original_Hamiltonian}), the value of which expresses the mass square of scalar field {$m_n^2 = |\omega|(n + 1/2)$} when we consider the higher-dimensional field theory on $M^4 \times T^2$ with magnetic flux.
In the cases of spinor and vector, there is a constant shift from the scalar case originating from their Lorentz structure and the explicit forms are {$m_n^2 = |\omega| n$ and $m_n^2 = |\omega|(n - 1/2)$}, respectively~{\cite{Cremades:2004wa,Hamada:2012wj}}.

\section{Operator formalism for $T^2/Z_N$ twisted orbifolds with magnetic flux
\label{section:T2/ZN}}

{
Based on the knowledge in the previous section, now we discuss the operator formalism describing magnetized $T^2 /Z_{N}$ twisted orbifold. We should emphasize {that} our results, {{\it e.g.}} the number of physical states, are consistent with the previous numerical results in Ref.~\cite{Abe:2013bca} and, moreover, our results are analytically exact. As we mentioned in Section~\ref{section:wavefunction-analysys}, it is quite nontrivial because of the following reasons.

In the wavefunction analysis, the physical states $f^{(j)}_{T^2/Z_{N}}(z)_{\eta}$ with a $Z_{N}$ eigenvalue $\eta$ on {the} magnetized $T^2 /Z_{N}$ twisted orbifold consist of {linear combinations} of $Z_{N}$-twisted zero-mode functions $f^{(j)}(\omega^k z)$ $ (k=0,1,2,\cdots , N-1)${:
	\begin{align}
	f^{(j)}_{T^2/Z_{N}}(z)_{\eta}= \mathcal{N}^{(j)}_{R,\eta}\sum_{x =0}^{N-1}\bar{\eta}^{x}\ f_{T^2}^{(j)}(\omega^x z),\label{physicalmodes}
	\end{align}
where} we omit the unimportant subscript. Since the $Z_{N}$-twisted mode function {$f_{T^2}^{(j)}(\omega^k z)$ satisfies the same equation as $f_{T^2}^{(j)}( z)$, $f^{(j)}_{T^{2}}(\omega^{x}z)$ has to be expanded in some linear combination of $f^{(j)}_{T^{2}}(z)$ as
	\begin{align}
	f^{(j)}_{T^2}(\omega^x z)= \sum_{m=0}^{|M|-1}C_{jm}^{(\omega^x)}f^{(m)}_{T^2}(z).
	\end{align}
If we can evaluate the coefficient $C_{jm}^{(\omega^x)}$} analytically, then we can obtain the results, {{\it e.g.},} the number of independent physical states, analytically from {Eq.~(\ref{RankK})}. However, it is not easy since the mode functions $f_{T^2}^{(j)}(z)$ contain the theta {functions (see Eq.~(\ref{fermion_mode_function}))}. 

Amazingly, in the operator analysis, we can evaluate the coefficient {$C_{jm}^{(\omega^x)}$} analytically as we will see in this section. Furthermore, the results of the operator formalism are available to not only the zero modes but also all KK modes. The correspondence between the numerical results in Ref.~\cite{Abe:2013bca} and the analytic results of ours {will be} found in the next section.
}

{The $T^2/Z_{N}$ twisted} orbifold enforces the discrete symmetry under $2\pi/N$-angle rotation, where it is well-known that only the $N=2,3,4,6$ are possible on $T^2$.
In the following part, the unitary operator $\widehat{U}_\theta$ manipulates the rotation with an angle $\theta$ around the {origin} $(y_1=y_2=0)$.
In the vector coordinate $\mathbf{y}$, the rotation is described by the two-by-two representation matrix $R_\theta$,
\al{
\mathbf{y} \to R_\theta \mathbf{y},\quad
R_\theta =
\begin{pmatrix} \cos{\theta} & -\sin{\theta} \\ \sin{\theta} & \cos{\theta} \end{pmatrix},
}
where the same expression is valid for the operators $\widehat{\mathbf{y}}$ and $\widehat{\mathbf{p}}$,
\al{
\widehat{\mathbf{y}} \to \widehat{U}_{\theta} \widehat{\mathbf{y}} \widehat{U}_{\theta}^{\dagger} = R_\theta \widehat{\mathbf{y}},\quad
\widehat{\mathbf{p}} \to \widehat{U}_{\theta} \widehat{\mathbf{p}} \widehat{U}_{\theta}^{\dagger} = R_\theta \widehat{\mathbf{p}}.
\label{rotation_operator_representation}
}

{
{
In the following discussion, we take the Wilson line phase {$\mathbf{a}$ as {zero:}
	\begin{align}
	\mathbf{a}={\bm 0}.
	\end{align}
It} has already been discussed in Ref.~\cite{Abe:2013bca} that we can remove the Wilson line phase by using a large gauge transformation without any loss of generality.
}
We can find the discussion on the issue in operator formalism in Appendix~\ref{appendix:gauge}.
By use of the results in Eq.~(\ref{rotation_operator_representation})}, {the} transformation of the Hamiltonian in Eq.~(\ref{system_operator}) under the rotation is evaluated as
\al{
\widehat{H} \to 
\widehat{U}_{\theta} \widehat{H} \widehat{U}_{\theta}^{\dagger} =
\left( \widehat{\mathbf{p}} {+} \frac{q}{2} R_{\theta}^{\mathrm{T}} \Omega R_{\theta} \widehat{\mathbf{y}} \right)^2,
}
and hence the following {condition is} required for invariance:
\al{
{\left[ R_{\theta}, \Omega \right] = 0.}
		\label{orbifoldcondition_on_Hamiltonian}
}
As shown in Eq.~(\ref{transformation_for_symmetricform}), only the unphysical symmetric part of $\Omega$ makes a nonzero contribution to the commutator $\left[ R_{\theta}, \Omega \right]$.
Therefore, {the condition} in Eq.~(\ref{orbifoldcondition_on_Hamiltonian}) is realized after the gauge fixing~(\ref{gauge_fixing}) irrespective of the value of $\theta$.
Then, we can conclude that the Hamiltonian itself is invariant under the rotation in spite of {the} value of the angle $\theta$.

{Although the Hamiltonian is invariant under the rotation $U_{\theta}$,} situations in the constraints on states {are found to be} nontrivial.
Remembering Eq.~(\ref{constraint_on_torus_states}), the two operators $\widehat{T}_a$ $(a=1,2)$ transform under the rotation generated by $\widehat{U}_{\theta}$ as
\al{
\widehat{T}_a \to \widehat{U}_{\theta} \widehat{T}_a \widehat{U}_{\theta}^{\dagger} =
({R_{\theta}^{\mathrm{T}}} \mathbf{u}_a)^{\mathrm{T}} \left( \widehat{\mathbf{p}} {+} \frac{q}{2} \Omega^{\mathrm{T}} \widehat{\mathbf{y}} \right),\quad
(a=1,2).
	\label{T_transformation}
}
Here, the vector ${R_{\theta}^{\mathrm{T}}} \mathbf{u}_a$ is {given by} a linear combination of $\mathbf{u}_1$ and $\mathbf{u}_2$, and subsequently, $\widehat{T}_1$ and $\widehat{T}_2$ are also mixed each other in general.
{It is convenient to choose the two basis vectors $\mathbf{u}_a$ in such a way that under} the rotations generated by $(R_{\theta=2\pi/N})^{\mathrm{T}}$, the two basis vectors are transformed as
{\al{
Z_2 \text{ case:}&\quad (R_{\theta=2\pi/N})^{\mathrm{T}} \mathbf{u}_1 = -\mathbf{u}_1,&
	(R_{\theta=2\pi/N})^{\mathrm{T}} \mathbf{u}_2 &= -\mathbf{u}_2, \notag \\
Z_3 \text{ case:}&\quad (R_{\theta=2\pi/N})^{\mathrm{T}} \mathbf{u}_1 = -\mathbf{u}_1 -\mathbf{u}_2,&
	(R_{\theta=2\pi/N})^{\mathrm{T}} \mathbf{u}_2 &= +\mathbf{u}_1, \notag \\
Z_4 \text{ case:}&\quad (R_{\theta=2\pi/N})^{\mathrm{T}} \mathbf{u}_1 = -\mathbf{u}_2,&
	(R_{\theta=2\pi/N})^{\mathrm{T}} \mathbf{u}_2 &= +\mathbf{u}_1, \notag \\
Z_6 \text{ case:}&\quad (R_{\theta=2\pi/N})^{\mathrm{T}} \mathbf{u}_1 = +\mathbf{u}_1 -\mathbf{u}_2,&
	(R_{\theta=2\pi/N})^{\mathrm{T}} \mathbf{u}_2 &= +\mathbf{u}_1{.}
	\label{basisvector_rotation}
}
Transformations of $\widehat{T}_a$ under the discrete rotations with angles {$\theta = 2\pi/N$} in $Z_N$ orbifoldings $(N=2,3,4,6)$ are easily evaluated by use of the results in Eq.~(\ref{basisvector_rotation})},
\al{
Z_2 \text{ case:}&\quad \widehat{T}_1 \to - \widehat{T}_1,\quad
	\widehat{T}_2 \to - \widehat{T}_2,
		\label{Trotation_Z2}\\
Z_3 \text{ case:}&\quad \widehat{T}_1 \to - \widehat{T}_1 - \widehat{T}_2,\quad
	\widehat{T}_2 \to \widehat{T}_1,
		\label{Trotation_Z3}\\
Z_4 \text{ case:}&\quad \widehat{T}_1 \to - \widehat{T}_2,\quad
	\widehat{T}_2 \to \widehat{T}_1,
		\label{Trotation_Z4}\\
Z_6 \text{ case:}&\quad \widehat{T}_1 \to \widehat{T}_1 - \widehat{T}_2,\quad
	\widehat{T}_2 \to \widehat{T}_1.
		\label{Trotation_Z6}
}

{To} investigate the number of physical states on the above magnetized $T^2 /Z_{N}$ orbifold, we will construct physical states via the following three steps:
\begin{enumerate}
\item Derivation of the allowed Scherk-Schwarz phases ($\alpha_{1},\alpha_{2}$)

\quad On $T^2/Z_{N}$, any $Z_{N}$-transformed state $\widehat{U}_{Z_{N}}|\psi\rangle$ should  satisfy the same constraint condition as $|\psi\rangle$, which is a state on $T^2$.
	\begin{align}
	e^{i \widehat{T}_a} \left( \widehat{U}_{Z_N} \ket{\psi} \right) = e^{2\pi i \alpha_a} \left( \widehat{U}_{Z_N} \ket{\psi} \right),\quad (a=1,2),\label{constraints_on_states_ZN}
	\end{align}
where we already fixed a gauge (\ref{gauge_fixing}) and we adopt a notation $\widehat{U}_{Z_N} := \widehat{U}_{2\pi/N}$. The above consistency condition will produce a restriction to the {values} of the Scherk-Schwarz phases ($\alpha_{1},\alpha_{2}$) as a consequence of the $Z_{N}$ orbifold. We {will} find that the number of allowed Scherk-Schwarz phases are {the same} as the number of fixed points on $T^2/Z_{N}$ orbifold. See Appendix~\ref{appendix:gauge}, {for} detail.

\item Derivation of the analytic form of a $Z_{N}$-transformed state $\left(\widehat{U}_{Z_{N}} \right)^x \Ket{\psi} $ ($x=0,1,2,\cdots,N-1$)

\quad Since the physical states $|\psi\rangle_{T^2/Z_{N}}$ on $T^2 /Z_{N}$ consist of {linear combinations} of the states $|\psi\rangle$ on $T^2$ with the projection operator $\widehat{\mathcal{P}}_{T^2/Z_N, {\eta}}$,
	\begin{align}
	&|\psi\rangle_{T^2 /Z_{N},\eta}=\widehat{\mathcal{P}}_{T^2/Z_N, {\eta}}|\psi\rangle,\\[0.2cm]
	&\widehat{\mathcal{P}}_{T^2/Z_N, {\eta}}:=\frac{1}{N}\left\{ \sum_{x=0}^{N-1} \bar{\eta}^{x} \left( \widehat{U}_{Z_N} \right)^x \right\},\label{defining_projection_operator}
	\end{align}
we need to obtain the analytic form of {the} $Z_{N}$-transformed state $\left(\widehat{U}_{Z_{N}} \right)^x \Ket{n, \frac{j}{M}} $. The {states} $\left|n, \frac{j}{M}\right\rangle$ and $\widehat{U}_{Z_{N}}\left|n, \frac{j}{M}\right\rangle$ satisfy the same equation with each other so that {$\left(\widehat{U}_{Z_{N}} \right)^x\left|n, \frac{j}{M}\right\rangle$ should} be expanded as follows:
	\begin{align}
	\left(\widehat{U}_{Z_{N}}\right)^{x}\left|n, \frac{j}{M}\right\rangle =\sum^{{|M|}-1}_{{m}=0}D^{(\omega^x)}_{jm}\left|n,  \frac{m}{M}\right\rangle. \label{ExpandCoefficient}
	\end{align}
To obtain the analytic form of $\left(\widehat{U}_{Z_{N}} \right)^x \Ket{n, \frac{j}{M}} $, we need to calculate the coefficient $D^{(\omega^x)}_{jm}$. However, as we mentioned in the beginning of this chapter, it is quite difficult to evaluate $D^{(\omega^x)}_{jm}$ in the wavefunction analysis. {It turns out that it can be evaluated} analytically and exactly by using operator formalism in this section.

\item Construction of the physical states

\quad With combining the above results, we can construct physical states {on $T^2/Z_{N}$}{,}
	\begin{align}
	\left|n,  \frac{j}{M}\right\rangle_{T^2/Z_{N},\eta}&=\widehat{\mathcal{P}}_{T^2/Z_N, {\eta}}\left|n, \frac{j}{M}\right\rangle,\nonumber\\[0.2cm]
	&=\frac{1}{N}\sum^{N-1}_{x=0}\bar{\eta}^{x} \sum^{{|M|}-1}_{k=0}D^{(\omega^x)}_{jk}\left|n,\frac{k}{M}\right\rangle.\label{OperatorPhysicalStates}
	\end{align}
\end{enumerate}

}
%
%
%
{Thus,} in the remaining part of this section, we first derive possible patterns of the two Scherk-Schwarz phases  {$(\alpha_1,\alpha_2)$ for every $Z_{N}$-orbifold ($N=2,3,4,6$) from Eq.~(\ref{constraints_on_states_ZN})}.
{We then calculate $D^{(\omega^{x})}_{jm}$ in (\ref{ExpandCoefficient}) and also the coefficients in (\ref{OperatorPhysicalStates}) in the case of N=2,3,4,6, separately.}
Finally, we construct the physical states on the magnetized $T^2/Z_{N}$ orbifolds and derive the number of physical states.



We emphasize that all the relations discussed in section~\ref{section:T2} should hold since the $Z_N$ discrete symmetry is additional on $T^2$.
As we showed in Eq.~(\ref{periodic_condition}), the system has the periodicity{,
\al{
\Ket{n, \widetilde{Y}} = \Ket{n, \widetilde{Y} -1},
}
and} the coordinate $\widetilde{Y}$ is quantized, where the possible values in Eq.~(\ref{coordinate_quantization_condition}) differentiate {the} degenerated states.
As {shown} in Eq.~(\ref{form_of_transration}), the operator $\Ptil$ works as the generator for the {translation} along $\widetilde{Y}$-direction.
The following expression is helpful,
{\al{
\Ket{n,\frac{j}{M}} = e^{-i \frac{j}{M} \Ptil} \ket{n,0}.
		\label{shift_operaton_on_state}
}}

Throughout the following part, the mathematical formula for arbitrary operators $\widehat{A}$ and $\widehat{B}$,
\al{
\widehat{A} e^{\widehat{B}} {\widehat{A}}^{-1} = e^{\widehat{A} \widehat{B} {\widehat{A}}^{-1}}
		\label{usefulformula1}
}
{is} very useful and we use the relations between the operators {$\widehat{T}_1$, $\widehat{T}_2$ and $\Ytil$, $\Ptil$} in Eq.~(\ref{new_operators}) many times.
Also, we comply with the three ``conventions":
\begin{itemize}
\item
We only consider the case $M>0$ {but we can analyze the case of $M<0$ in the same way}.
\item
Fundamental regions of $\alpha_1$, $\alpha_2$ and $\widetilde{Y}$ (in the eigenstate of $\widehat{\widetilde{Y}}$) are selected as $[0,1)$ (without lose of generality).
\item
We omit to describe eigenvalues of the energy since the following discussion is viable in every energy eigenstate{,}
	\al{
	\left| n,\frac{j}{M}\right\rangle \rightarrow \left| \frac{j}{M}\right\rangle{.}
	}
\end{itemize}

\subsection{$T^2/Z_2$}

{
First, we calculate the allowed Scherk-Schwarz phases {on $T^2/Z_{2}$} by using Eq.~(\ref{constraints_on_states_ZN}).
}This case is not complicated since the two operators $\widehat{T}_1$ and $\widehat{T}_2$ are not mixed under the operation of $\Utwo$, where only their signs are flipped.
After requesting {the} coexistence of the conditions {in Eqs.~(\ref{constraints_on_states})} and (\ref{constraints_on_states_ZN}), we obtain the requirements,
\al{
\alpha_1 = -\alpha_1 \quad \text{and} \quad \alpha_2 = -\alpha_2 \quad (\text{mod }1),
}
where we {used} the relation in Eqs.~(\ref{Trotation_Z2}) and (\ref{usefulformula1}).
As {shown in~\cite{Abe:2013bca}}, the two phases {can take} individual values and the possibilities are
\al{
(\alpha_1, \alpha_2) =
\left(0,0 \right),
\left(\frac{1}{2}, 0 \right),
\left(0,\frac{1}{2} \right),
\left(\frac{1}{2}, \frac{1}{2} \right).
		\label{allowedphases_Z2}
}

Next, we consider {$Z_{2}$-transformed states.}
Here, we remember the statement in section~\ref{section:T2} for the state on $T^2$ that degenerated states of an energy can be {specified} by the {eigenvalues} of the operator $\widehat{\widetilde{Y}}$.
Thereby, we first try to find a state which includes the operator $\Utwo$ and is also an eigenstate of $\widehat{\widetilde{Y}}$.
We {find such a state} as $\Utwo \ket{j/M}$ with the eigenvalue {$(-2\alpha_1 - j)/M$} from the following calculation{,
\al{
e^{2\pi i \Ytil} \left( \Utwo \Ket{\frac{j}{M}} \right) &=\Utwo\left(\Utwo^{\dagger} e^{2\pi i \left(\frac{\widehat{T}_{1}}{2\pi M}-\frac{\alpha_{1}}{M}\right)}\Utwo\right)\Ket{\frac{j}{M}}\nonumber\\
&=\Utwo\, e^{2\pi i \left(-\frac{\widehat{T}_{1}}{2\pi M}-\frac{\alpha_{1}}{M}\right)}\Ket{\frac{j}{M}}\nonumber\\
&=\Utwo\, e^{-2\pi i \Ytil-2\pi i \frac{2\alpha_{1}}{M}}\Ket{\frac{j}{M}}\nonumber\\
&= e^{2\pi i \left(\frac{{-} 2\alpha_1 - j}{M}\right)} \left( \Utwo \Ket{\frac{j}{M}} \right),
}
and} then we can write down the state $\Utwo \ket{j/M}$ as follows:
\al{
\Utwo \Ket{\frac{j}{M}} = e^{2 \pi i \eta_j} \Ket{\frac{- 2\alpha_1 - j}{M}},
		\label{condition1_Z2}
}
{where} $\eta_j$ expresses {a phase ambiguity to be determined}.\footnote{
Here, the norm should not be changed under the rotation.
}
{
Since there {exists} the relation on the translation along $\widetilde{Y}$-direction in Eq.~(\ref{shift_operaton_on_state}), we have another expression for the above equation:
	\begin{align}
	\Utwo \Ket{\frac{j}{M}} = \Utwo e^{-i \frac{j}{M} \Ptil} \ket{0} =
{\left( \Utwo e^{-i \frac{j}{M} \Ptil} \Utwo^\dagger \right) \Utwo \ket{0}} =
e^{2\pi i \left( -\frac{j}{M} 2\alpha_2 + \eta_0 \right)} \Ket{\frac{- 2\alpha_1 - j}{M}},
		\label{condition2_Z2}
	\end{align}
where we {used} the relations in {Eqs.~(\ref{new_operators}),} (\ref{Trotation_Z2}), (\ref{shift_operaton_on_state}) and (\ref{condition1_Z2}). Comparing the two equations (\ref{condition1_Z2}) and (\ref{condition2_Z2}), we have the relation
	\begin{align}
	\eta_j = \eta_0 - \frac{j}{M} 2\alpha_2 \quad (\text{mod } 1).
	\label{Z2_resultform1}
	\end{align}
}

The $Z_2$-consistency requires the condition $\Utwo^2 = \widehat{1}$ for every state, where $\widehat{1}$ is the unit operator.
This fact leads to the condition,
\al{
\Utwo^2 \Ket{\frac{j}{M}}  = e^{2\pi i (\eta_j + \eta_{- 2\alpha_1 -j})} \Ket{\frac{j}{M}} 
= \Ket{\frac{j}{M}},
}
which is equivalent to
{
\al{
\eta_j + \eta_{- 2\alpha_1 -j} = 0 \quad (\text{mod }1). \label{U2consistency}
}}

{Combining the results in {Eqs.~(\ref{allowedphases_Z2}), (\ref{Z2_resultform1})} and (\ref{U2consistency}), }we can determine the form of $\eta_j$ as
\al{
\eta_j = -\frac{2\alpha_2}{M}  (j + \alpha_1),
\label{Z2_twistphase_finalform}
}
where we omit the trivial overall phase. 
{The explicit analytic form of the $Z_{2}$-transformed state $\widehat{U}_{Z_{2}}\left|\frac{j}{M}\right\rangle$, which corresponds to Eq.~(\ref{ExpandCoefficient}), is the {following:}
	\begin{align}
	\widehat{U}_{Z_{2}}\left| \frac{j}{M}\right\rangle=e^{-2\pi i \cdot\frac{2\alpha_{2}}{M}(j+\alpha_{1})}\left| \frac{-2\alpha_{{1}}-j}{M}\right\rangle, \label{T^2/Z_2}
	\end{align}
which is equivalent to the following expression{:}
	\begin{align}
	&\widehat{U}_{Z_{2}}\left| \frac{j}{M}\right\rangle=\sum_{k=0}^{M-1}D^{(\omega)}_{jk}\left|\frac{k}{M}\right\rangle{,}\nonumber\\[0.15cm]
	&D^{(\omega)}_{jk}=e^{-2\pi i \cdot\frac{2\alpha_{2}}{M}(j+\alpha_{1})}\delta_{-2{\alpha_{1}}-j,k}. \label{T^2/Z_2part2}
	\end{align}
Finally, we can construct the physical states on $T^2/Z_{2}$ by {using Eq.~(\ref{OperatorPhysicalStates}){,}}
	\begin{align}
	\left|\frac{j}{M}\right\rangle_{T^2/Z_{2},\eta}&=\frac{1}{2}\sum^{1}_{x=0}\bar{\eta}^{x}\sum_{k=0}^{M-1}D_{jk}^{(\omega^{x})}\left|\frac{k}{M}\right\rangle\nonumber\\
	&=:\sum_{k=0}^{M-1}M_{jk}^{(Z_{2};\eta)}\left|\frac{k}{M}\right\rangle,\hspace{2em}(\eta=+1,-1).
	\end{align}
The number of independent physical states $\left|\frac{j}{M}\right\rangle_{T^2/Z_{2},\eta}$ is nothing but the rank of the $M$-by-$M$ matrix $M_{jk}^{(Z_{2};\eta)}${,}
	\begin{align}
	\text{The number of physical states }\left|n,\frac{j}{M}\right\rangle_{T^2/Z_{2},\eta}= \text{Rank}\left[M_{jk}^{(Z_{2};\eta)}\right].
	\end{align}
{The analytic results for the $T^2/Z_{2}$ {have already been} given by Ref.~\cite{Abe:2013bca}.}
}

\subsection{$T^2/Z_3$}

First, we note that in the cases of $T^2/Z_3$ and the following $T^2/Z_6$, {a linear combination of the two operators $\widehat{T}_a$ $(a=1,2)$ appears to a variable of the exponential} and the  Baker-Campbell-Hausdorff formula (or the Zassenhaus formula) for {two operators $\widehat{A}$ and $\widehat{B}$ with $t \in \mathbb{R}$ is helpful:
\al{
e^{t (\widehat{A} + \widehat{B})} = e^{t \widehat{A}} e^{t \widehat{B}} e^{-\frac{t^2}{2} \left[ \widehat{A}, \widehat{B} \right]} \quad
\left(\text{when } \left[ \widehat{A}, \left[ \widehat{A}, \widehat{B} \right] \right] = 0
\text{ and } \left[ \widehat{B}, \left[ \widehat{A}, \widehat{B} \right] \right] = 0 \right).
		\label{BCH_formula}
}
Since the commutation relation} associated with $\widehat{T}_1$ and $\widehat{T}_2$ {is given by
\al{
\left[ \widehat{T}_1, \widehat{T}_2 \right] = {-} 2\pi i M,
}
the} precondition in Eq.~(\ref{BCH_formula}) is trivially fulfilled for $\widehat{T}_a$s.

{Now, to obtain the allowed {Scherk}-Schwarz phases, we follow the same strategy {as} in the case of $T^2/Z_2$. Then, from Eq.~(\ref{constraints_on_states_ZN}),} $\alpha_a$s should satisfy the conditions
\al{
\alpha_1 = \alpha_2 \quad \text{and} \quad
{\alpha_2 = -\alpha_2 -\alpha_1 {-} \frac{M}{2}} \quad (\text{mod }1),
}
for {$T^2/Z_{3}$} and the resultant allowed combinations are as follows:
{
\al{
\alpha &:= \alpha_1 = \alpha_2, \label{Z3alpha}\\[0,2cm]
\alpha&=
	\begin{cases}
	\displaystyle 0,\, \frac{1}{3},\, \frac{2}{3} & \text{for } M : \text{even}, \\[4mm]
	\displaystyle \frac{1}{6},\, \frac{3}{6},\, \frac{5}{6} & \text{for } M : \text{odd}. 
	\end{cases}
}
Interestingly, when the value of $M$ is odd, the {$Z_3$-orbifold} system cannot be defined without the Scherk-Schwarz phases.

{To evaluate the form {of} the $Z_{3}$-transformed state $\Uthree \ket{j/M}$ , we act the operator $e^{2\pi i \Ytil}$ to it. In the case of $T^2/Z_3$}, the state $\Uthree \ket{j/M}$ {is not an eigenstate of} $\Ytil$ {because the argument of the ket vector gets to be different from the original one {as $j/M \to (j+1)/M$ after manipulating {$e^{ 2\pi i \Ytil}$} to the $Z_{3}$-rotated state $\Uthree \ket{j/M}$}:
\al{
e^{2\pi i \Ytil} \Uthree \Ket{\frac{j}{M}} &= \Uthree\left(\Uthree^{\dagger} e^{2\pi i \left(\frac{\widehat{T}_{1}}{2\pi M}-\frac{\alpha_{1}}{M}\right)}\Uthree\right)\Ket{\frac{j}{M}}\nonumber\\
&=\Uthree\,e^{2\pi i \left(\frac{\widehat{T}_{2}}{2\pi M}-\frac{\alpha_{1}}{M}\right)}\Ket{\frac{j}{M}}\nonumber\\
&=\Uthree\,e^{2\pi i \left(-\frac{\Ptil}{2\pi M}+\frac{\alpha_{2}}{M}-\frac{\alpha_{1}}{M}\right)}\Ket{\frac{j}{M}}\nonumber\\
&=\Uthree\,e^{-i\frac{\Ptil}{M}}\Ket{\frac{j}{M}}\nonumber\\
&= {\Uthree} \Ket{\frac{j+1}{M}},
} 
where we used Eq.~(\ref{Z3alpha}).}
This fact gives us a hint {for finding a suitable form of an eigenstate of $\Ytil$}.
Effects of this shift would be cancelled out after taking summation over the index $j$.
Based on this speculation, we focus on the following state,
\al{
\sum_{j=0}^{M-1} \Uthree \Ket{\frac{j}{M}}.
		\label{Z3_relation1}
}
{Since} the above state satisfies the condition,
	\begin{align}
	e^{ 2\pi i \Ytil}\sum_{j=0}^{M-1} \Uthree \Ket{\frac{j}{M}}=\sum_{j=0}^{M-1} \Uthree \Ket{\frac{j}{M}},
	\end{align}
%
the following representation {should be completed} with two parameters $\A$ and $\theta$, which show the magnitude and phase parts of the undetermined coefficient{{,}
	\begin{align}
	\sum_{j=0}^{M-1}  \Uthree \Ket{\frac{j}{M}}= \A e^{i\theta} \Ket{0}.
	\end{align}
	
With} casting the shift operator $e^{-i\frac{l}{M}\widehat{\widetilde{P}}}$ on both sides, we {obtain
\al{
 e^{-i\frac{\pi}{M}l^2- i \frac{6\pi\alpha}{M}l} \sum_{j=0}^{M-1}e^{-2\pi i \frac{l\cdot j}{M}}\Uthree \Ket{\frac{j}{M}} = \A e^{i\theta} \Ket{\frac{l}{M}},
		\label{Z3_sumtosingle}
}
and} subsequently, we can derive the following simple form from the above with summing over $l$ from $0$ to $M-1${,
\al{
\Uthree \Ket{0} = \frac{\A}{M} e^{i\theta} \sum_{l=0}^{M-1} e^{i\frac{\pi}{M}l^2 +i\frac{6\pi\alpha}{M}l}\Ket{\frac{l}{M}}. \label{U30}
}
Here,} we used the formula,
\al{
\sum_{k=0}^{M-1} e^{2\pi i \frac{s}{M} k} = M \delta_{s,0}\, ,\quad
{(s=0,1,\cdots,M-1),}
		\label{root_formula}
}
which is proved with ease via properties of $M$-th root of {unity}.
}

{The} $Z_3$-consistency $(\Uthree)^3 = \widehat{1}$ pins down values of $\A$ and $\theta$ immediately when we utilize the formula{,
\al{
\sum_{s=0}^{M-1} e^{-\pi i \frac{(s + t {\pm} \beta)^2}{M}} = \sqrt{M} e^{-\frac{1}{4} \pi i}
\quad
\text{for } t \in \mathbb{Z},\ \beta =
\begin{cases} \displaystyle 0 & \text{for } M \text{: even}, \\[2mm]
\displaystyle \frac{1}{2} & \text{for } M \text{: odd}. \end{cases}
		\label{magic_formula}
} 
We} mention that {the} above summation takes the universal form irrespective of the choice of {$t$, $\beta$ and the sign in front of $\beta$} (within the shown ranges in Eq.~(\ref{magic_formula})).
The derivation of this is provided in {Appendix~\ref{appendix:derivation}}. After some calculations, resultant values are declared, ({{\it e.g.,}} via $\Uthree^3 \ket{0} = \ket{0}$) as
\al{
\A = \sqrt{M}, \quad \theta = {-} \frac{\pi}{12} {+} \frac{3\pi}{M}\alpha^2,
}
where we {ignored} the trivial overall phase in $\theta$.

Now, we can obtain the form of the $Z_{3}$-transformed state $\Uthree \Ket{\frac{j}{M}} $ for the projection operator from Eq.~(\ref{U30}) with the shift operation in Eq.~(\ref{shift_operaton_on_state}) {as}
\al{
\Uthree \Ket{\frac{j}{M}} = \left(\Uthree^\dagger\right)^2 \Ket{\frac{j}{M}} &=
	{\frac{1}{\sqrt{M}} e^{-i\frac{\pi}{12} + i\frac{3\pi\alpha^2}{M} }
	\sum_{k=0}^{M-1} e^{i\frac{\pi}{M}k(k+6\alpha)+2\pi i \frac{j\cdot k}{M}} \Ket{\frac{k}{M}}}, \notag \\
\left(\Uthree\right)^2 \Ket{\frac{j}{M}} = \Uthree^\dagger \Ket{\frac{j}{M}} &=
	{\frac{1}{\sqrt{M}} e^{i\frac{\pi}{12} - i\frac{3\pi\alpha^2}{M}-i\frac{\pi}{M}j(j+6\alpha)}
	\sum_{k=0}^{M-1} e^{-2\pi i \frac{j\cdot k}{M}} \Ket{\frac{k}{M}}},
	\label{finalresults_T2Z3}
}
which {are} equivalent to the following expressions:
	\begin{align}
	&\left(\widehat{U}_{Z_{3}}\right)^{x}\left|\frac{j}{M}\right\rangle=\sum_{k=0}^{{M}-1}D^{(\omega^x)}_{jk}\left|\frac{k}{M}\right\rangle,\hspace{3em}(x=0,1,2),\nonumber\\
	&D^{(\omega)}_{jk}=\frac{1}{\sqrt{M}} e^{-i\frac{\pi}{12} + i\frac{3\pi\alpha^2}{M} }e^{i\frac{\pi}{M}k(k+6\alpha)+2\pi i \frac{j\cdot k}{M}},\nonumber\\
	&D^{(\omega^2)}_{jk}=\frac{1}{\sqrt{M}} e^{i\frac{\pi}{12} - i\frac{3\pi\alpha^2}{M}-i\frac{\pi}{M}j(j+6\alpha)}e^{-2\pi i \frac{j\cdot k}{M}}.\label{finalresults_T2Z3part2}
	\end{align}
{We again comment that the above analytic results of the $Z_{3}$-transformed state are nontrivial. In the case of the {wavefunction} analysis, we need to rely on {numerical calculations} since the states are given by theta functions. On the other hand, we can evaluate the exact form of the $Z_{3}$-transformed {states} in this case and, moreover, the above results are applicable to all KK-modes since they are irrelevant to the principal quantum number $n$ of the state $\left| n,\frac{j}{M}\right\rangle$. We can say the same thing for the following $T^2 /Z_4$ and $T^2/Z_6$ cases.}

Finally, we construct physical states on $T^2/Z_{3}$. By using {Eq.~(\ref{OperatorPhysicalStates})}, the physical states are represented {by}
	\begin{align}
	\left|\frac{j}{M}\right\rangle_{T^2/Z_{3},\eta}&=\frac{1}{3}\sum_{x=0}^{2}\bar{\eta}^{x}\sum_{k=0}^{M-1}D^{(\omega^x)}_{jk}\left|\frac{k}{M}\right\rangle\nonumber\\
	&=:\sum^{M-1}_{k=0}M_{jk}^{(Z_{3};\eta)}\left|\frac{k}{M}\right\rangle,\hspace{4em}(\eta=1,\omega,\omega^2),
	\end{align}
where
	\begin{align}
	M_{jk}^{(Z_{3};\eta)}=\frac{1}{3}\sum_{x=0}^{2}\bar{\eta}^{x}D^{(\omega^x)}_{jk}.
	\end{align}

The number of independent physical states $\left|\frac{j}{M}\right\rangle_{T^2/Z_{3},\eta}$ is nothing but the rank of the matrix $M_{jk}^{(Z_{3};\eta)}${,}
	\begin{align}
	\text{The number of physical states }\left|\frac{j}{M}\right\rangle_{T^2/Z_{3},\eta}= \text{Rank}\left[M_{jk}^{(Z_{3};\eta)}\right].
	\end{align}
After investigating the rank of the matrix $M_{jk}^{(Z_{3};\eta)}$, we obtain the results shown in {Table~\ref{T2Z3Psi_1e}, \ref{T2Z3Psi_2e}, \ref{T2Z3Psi_1o} and \ref{T2Z3Psi_2o}}. We, again, emphasize that the following results are completely consistent with the previous wavefunction analysis {\cite{Abe:2013bca}}.
{The correspondence between the matrix $M_{jk}^{(Z_{3};\eta)}$ and the kinetic matrix {$\mathcal{K}_{jk}^{(Z_{3};\eta)}$} will be discussed in Section~\ref{section:analysis} .}
\begin{table}[H]
\vspace{8mm}
\begin{center}
\begin{tabular}{|c|c|ccccccc|}
\hline
\multicolumn{2}{|c|}{$|M|$}& 2 & 4 & 6 & 8 & 10 & 12 & 14 \\
\hline
\multirow{3}{*}{$\eta$}&1 & 1 & 1 & 3 & 3 & 3 & 5 & 5 \\
&$\omega$ & 0 & 2 & 2 & 2 & 4 & 4 & 4 \\
&$\bar{\omega}$ & 1 & 1 & 1 & 3 & 3 & 3 & 5 \\
\hline
\end{tabular}
\end{center}
\vspace{-3mm}
\caption{The {numbers} of linearly independent zero-mode eigenstates with $Z_3$ eigenvalue $\eta$
  for $M=\mathrm{even}$ and
  $(\alpha_1,\alpha_2)=(0, 0)$ 
on $T^2/Z_3$.}
\label{T2Z3Psi_1e}
\end{table}
\begin{table}[H]
\vspace{12mm}
\begin{center}
\begin{tabular}{|c|c|ccccccc|}
\hline
\multicolumn{2}{|c|}{$|M|$}& 2 & 4 & 6 & 8 & 10 & 12 & 14 \\
\hline
\multirow{3}{*}{$\eta$}&1 & 1 & 2 & 2 & 3 & 4 & 4 & 5 \\
&$\omega$ & 1 & 1 & 2 & 3 & 3 & 4 & 5 \\
&$\bar{\omega}$ & 0 & 1 & 2 & 2 & 3 & 4 & 4 \\
\hline
\end{tabular}
\end{center}
\vspace{-3mm}
\caption{The {numbers} of linearly independent zero-mode eigenstates with $Z_3$ eigenvalue $\eta$
  for $M=\mathrm{even}$ and
  $(\alpha_1,\alpha_2)=\left({1\over 3},{1\over
      3}\right),\left({2\over 3},{2\over 3}\right)$ 
on $T^2/Z_3$.}
\label{T2Z3Psi_2e}
\end{table}
\begin{table}[H]
\vspace{12mm}
\begin{center}
\begin{tabular}{|c|c|ccccccc|}
\hline
\multicolumn{2}{|c|}{$|M|$}& 1 & 3 & 5 & 7 & 9 & 11 & 13 \\
\hline
\multirow{3}{*}{$\eta$}&1 & 1 & 1 & 2 & 3 & 3 & 4 & 5 \\
&$\omega$ & 0 & 1 & 2 & 2 & 3 & 4 & 4 \\
&$\bar{\omega}$ & 0 & 1 & 1 & 2 & 3 & 3 & 4 \\
\hline
\end{tabular}
\end{center}
\vspace{-3mm}
\caption{The {numbers} of linearly independent zero-mode eigenstates with $Z_3$ eigenvalue $\eta$
  for $M=\mathrm{odd}$ and
  $(\alpha_1,\alpha_2)=\left({1\over 6},{1\over
      6}\right),\left({5\over 6},{5\over 6}\right)$ 
on $T^2/Z_3$.}
\label{T2Z3Psi_1o}
\end{table}
\begin{table}[H]
\vspace{12mm}
\begin{center}
\begin{tabular}{|c|c|ccccccc|}
\hline
\multicolumn{2}{|c|}{$|M|$}& 1 & 3 & 5 & 7 & 9 & 11 & 13 \\
\hline
\multirow{3}{*}{$\eta$}&1 & 0 & 2 & 2 & 2 & 4 & 4 & 4 \\
&$\omega$ & 1 & 1 & 1 & 3 & 3 & 3 & 5 \\
&$\bar{\omega}$ & 0 & 0 & 2 & 2 & 2 & 4 & 4 \\
\hline
\end{tabular}
\end{center}
\vspace{-3mm}
\caption{The {numbers} of linearly independent zero-mode eigenstates with $Z_3$ eigenvalue $\eta$
  for $M=\mathrm{odd}$ and
  {$(\alpha_1,\alpha_2)=\left({3\over 6},{3\over 6}\right)$}
on $T^2/Z_3$.}
\label{T2Z3Psi_2o}
\end{table}


\subsection{$T^2/Z_4$}

{In the case of $T^2/Z_4$, the situation is similar to $T^2 /Z_{2}$ case but is somewhat complicated.
The way for determining allowed {values} of $\alpha_1$ and $\alpha_2$ is the same as in $T^2/Z_2$ and $T^2/Z_{3}$. By using Eqs.~(\ref{Trotation_Z4}), {(\ref{constraints_on_states_ZN})} and (\ref{usefulformula1}), we find that the following two conditions should be {fulfilled}:}
\al{
\alpha_1 = \alpha_2 \quad \text{and} \quad \alpha_1 = - \alpha_2 \quad (\text{mod }1).
}
Here, we conclude that the consistent {values} are
\al{
&\alpha := \alpha_1 = \alpha_2 \\
&\alpha= 0\ {\rm or}\  \frac{1}{2},
}
which {are} a subset of the result in $T^2/Z_2$ since the $Z_4$ orbifolding includes the $Z_2$ operation.

{To evaluate the form of the $Z_{4}$-transformed state $\widehat{U}_{Z_{4}}| j/M\rangle$, we act the operator $e^{ 2\pi i \Ytil}$ to the transformed state.}
{Unlike} $T^2/Z_2$, the state $\Ufour \ket{j/M}$ itself is not an eigenstate of $\Ytil$ as {in the case of} the $T^2/Z_{3}${, {\it i.e.},}
\al{
e^{ 2\pi i \Ytil} \left(\Ufour \Ket{\frac{j}{M}}\right) &=\Ufour \left(\Ufour^{\dagger}e^{ 2\pi i \left(\frac{\widehat{T}_{1}}{2\pi M}-\frac{\alpha_{1}}{M}\right)} \Ufour \right)\Ket{\frac{j}{M}}\nonumber\\
&=\Ufour\, e^{ 2\pi i \left(\frac{\widehat{T}_{2}}{2\pi M}-\frac{\alpha_{1}}{M}\right)}\Ket{\frac{j}{M}}\nonumber\\
&=\Ufour\, e^{ 2\pi i \left(-\frac{\Ptil}{2\pi M}+\frac{\alpha_{2}}{M}-\frac{\alpha_{1}}{M}\right)}\Ket{\frac{j}{M}}\nonumber\\
&=\Ufour\, e^{ -i\frac{\Ptil}{M}}\Ket{\frac{j}{M}}\nonumber\\
&=  \Ufour \Ket{\frac{j+1}{M}}.
		\label{Z4_relation1}
}
Then we again consider the following state as a candidate for {$\Ytil$-eigenstates} as $T^2/Z_3$,
\al{
\sum_{j=0}^{M-1}  \Ufour \Ket{\frac{j}{M}}.
}
{Using} Eq.~(\ref{Z4_relation1}), the following condition holds,
\al{
{e^{2\pi i \Ytil}}
\sum_{j=0}^{M-1}  \Ufour \Ket{\frac{j}{M}}
=
\sum_{j=0}^{M-1}  \Ufour \Ket{\frac{j}{M}}.
}
{Then} it should be rewritten {as
\al{
\sum_{j=0}^{M-1}\Ufour \Ket{\frac{j}{M}} = \A e^{i\theta} \Ket{0},
}
and} casting the operator $e^{-i \frac{l}{M} \Ptil}$ on {both sides} brings us {to
\al{
e^{-i\frac{4\pi\alpha}{M}l}\sum_{j=0}^{M-1} e^{-2\pi i \frac{l\cdot j}{M}} \,\Ufour \Ket{\frac{j}{M}} = \A e^{i\theta} \Ket{\frac{l}{M}}.
		\label{Z4_relation3}
}
The} parameters $\A$ and $\theta$ show the magnitude and phase parts of the undetermined coefficient{, respectively} as in the case of $T^2/Z_{3}$.
Here, taking summation over $l$ from $0$ to $M-1$ in Eq.~(\ref{Z4_relation3}) leads to the simple relation,\
\al{
\Ufour \Ket{0} = \frac{\A}{M} e^{i\theta} \sum_{l=0}^{M-1} e^{i\frac{4\pi \alpha}{M}l}\Ket{\frac{l}{M}}, \label{U40}
}
where we used the formula Eq.~(\ref{root_formula}).
The form (\ref{U40}) is just a part of what we would like to obtain.

{The values} of $\A$ and $\theta$ {can be} derived when we examine the consistency condition coming from $Z_4$ symmetry,  $\Ufour^4 = \widehat{1}$.
After making use of Eq.~(\ref{root_formula}), we can reach the result,\footnote{
The relation $\Ufour^4 = \widehat{1}$ is valid irrespective of operated states.
The easiest way to determine $\A$ and $\theta$ is to use the state $\ket{0}$ ($\Ufour^4 \ket{0} = \ket{0}$).
}
\al{
\A = \sqrt{M}, \quad {\theta = \frac{2\pi}{M} \alpha^2},
}
where we {dropped} the overall phase in $\theta$.

Now, {the form of the $Z_{4}$-transformed state $\Ufour \ket{j/M}$} is easily evaluated {from Eq.~(\ref{U40}) by using }the shift operation in Eq.~(\ref{shift_operaton_on_state}).
Here, we summarize the results for constructing the projective operator,
\al{
\Ufour \Ket{\frac{j}{M}} &= \left(\Ufour^\dagger\right)^3 \Ket{\frac{j}{M}} =
	{\frac{1}{\sqrt{M}} e^{2\pi i \frac{\alpha^2}{M}}
	\sum_{k=0}^{M-1} e^{2\pi i\frac{2\alpha}{M}k +2\pi i \frac{j\cdot k}{M}} \Ket{\frac{k}{M}}}, \notag \\
\left(\Ufour\right)^2 \Ket{\frac{j}{M}} &= \left(\Ufour^\dagger\right)^2 \Ket{\frac{j}{M}} =
	{e^{-2\pi i \frac{2\alpha}{M}(\alpha+j)} \Ket{\frac{-2\alpha - j}{M}}}, \notag \\
\left(\Ufour\right)^3 \Ket{\frac{j}{M}} &= \Ufour^\dagger \Ket{\frac{j}{M}} =
	{\frac{1}{\sqrt{M}} e^{-2\pi i \frac{\alpha^2}{M}-2\pi i \frac{2\alpha}{M}j }
	\sum_{k=0}^{M-1} e^{-2\pi i \frac{j\cdot k}{M}} \Ket{\frac{k}{M}}},
	\label{finalresults_T2Z4}
}
which {are} equivalent to the following expressions:
	\begin{align}
	&\left(\widehat{U}_{Z_{4}}\right)^{x}{\left|\frac{j}{M}\right\rangle}=\sum_{k=0}^{M-1}D_{jk}^{(\omega^x)}\left|\frac{k}{M}\right\rangle,\hspace{3em}(x=0,1,2,3),\nonumber\\
	&D_{jk}^{(\omega)}=\frac{1}{\sqrt{M}} e^{2\pi i \frac{\alpha^2}{M}} e^{2\pi i \frac{j\cdot k}{M}+2\pi i\frac{2\alpha}{M}k },\nonumber\\
	&D_{jk}^{(\omega^2)}=e^{-2\pi i \frac{2\alpha}{M}(\alpha+j)} \delta_{{-2\alpha-j,k}},\nonumber\\
	&D_{jk}^{(\omega^3)}=\frac{1}{\sqrt{M}} e^{-2\pi i \frac{\alpha^2}{M}-2\pi i \frac{2\alpha}{M}j }e^{-2\pi i \frac{j\cdot k}{M}}. \label{finalresults_T2Z4part2}
	\end{align}

Finally, we construct physical states on $T^2/Z_{4}$ by using {Eq.~(\ref{OperatorPhysicalStates})}{,}
	\begin{align}
	\left|\frac{j}{M}\right\rangle_{T^2/Z_{4},\eta}&=\frac{1}{4}\sum_{x=0}^{3}\bar{\eta}^{x}\sum_{k=0}^{M-1}D^{(\omega^x)}_{jk}\left|\frac{k}{M}\right\rangle\nonumber\\
	&=:\sum^{M-1}_{k=0}M_{jk}^{(Z_{4};\eta)}\left|\frac{k}{M}\right\rangle,\hspace{4em}(\eta=1,\omega,\omega^2,\omega^3),
	\end{align}
where
	\begin{align}
	M_{jk}^{(Z_{4};\eta)}=\frac{1}{4}\sum_{x=0}^{3}\bar{\eta}^{x}D^{(\omega^x)}_{jk}.
	\end{align}
The number of independent physical states $\left|\frac{j}{M}\right\rangle_{T^2/Z_{4},\eta}$ is nothing but the rank of the matrix $M_{jk}^{(Z_{4};\eta)}${,}
	\begin{align}
	\text{The number of physical states }\left|n,\frac{j}{M}\right\rangle_{T^2/Z_{4},\eta}= \text{Rank}\left[M_{jk}^{(Z_{4};\eta)}\right].
	\end{align}
After investigating the rank of the matrix $M_{jk}^{(Z_{4};\eta)}$, we obtain {the results shown in Table~\ref{T2Z4Psi_1} and \ref{T2Z4Psi_2}}. We emphasize that the following results are consistent with the previous wavefunction analysis {\cite{Abe:2013bca}}. 
\begin{table}[H]
\vspace{4mm}
\begin{center}
\begin{tabular}{|c|c|ccccccccccccccccc|}
\hline
\multicolumn{2}{|c|}{$|M|$}& 1 & 2 & 3 & 4 & 5 & 6 & 7 & 8 & 9 & 10 & 11 & 12 & 13 & 14 & 15 & 16 & 17  \\
\hline
\multirow{4}{*}{$\eta$}&$+1$ & 1 & 1 & 1 & 2 & 2 & 2 & 2 & 3 & 3 & 3 & 3 & 4 & 4 & 4 & 4 & 5 & 5 \\
&$+i$ & 0 & 0 & 1 & 1 & 1 & 1 & 2 & 2 & 2 & 2 & 3 & 3 & 3 & 3 & 4 & 4 & 4 \\
&$-1$ & 0 & 1 & 1 & 1 & 1 & 2 & 2 & 2 & 2 & 3 & 3 & 3 & 3 & 4 & 4 & 4 & 4 \\
&$-i$ & 0 & 0 & 0 & 0 & 1 & 1 & 1 & 1 & 2 & 2 & 2 & 2 & 3 & 3 & 3 & 3 & 4 \\
\hline
\end{tabular}
\end{center}
\vspace{-3mm}
\caption{The {numbers} of linearly independent zero-mode eigenstates with $Z_4$ eigenvalue $\eta$
for $(\alpha_1,\alpha_2)=(0, 0)$ on $T^2/Z_4$.}
\label{T2Z4Psi_1}
\end{table}
\begin{table}[H]
\vspace{6mm}
\begin{center}
\begin{tabular}{|c|c|cccccccccccccccc|}
\hline
\multicolumn{2}{|c|}{$|M|$}& 1 & 2 & 3 & 4 & 5 & 6 & 7 & 8 & 9 & 10 & 11 & 12 & 13 & 14 & 15 & 16  \\
\hline
\multirow{4}{*}{$\eta$}&$+1$ & 0 & 1 & 1 & 1 & 1 & 2 & 2 & 2 & 2 & 3 & 3 & 3 & 3 & 4 & 4 & 4 \\
&$+i$ & 1 & 1 & 1 & 1 & 2 & 2 & 2 & 2 & 3 & 3 & 3 & 3 & 4 & 4 & 4 & 4 \\
&$-1$ & 0 & 0 & 0 & 1 & 1 & 1 & 1 & 2 & 2 & 2 & 2 & 3 & 3 & 3 & 3 & 4 \\
&$-i$ & 0 & 0 & 1 & 1 & 1 & 1 & 2 & 2 & 2 & 2 & 3 & 3 & 3 & 3 & 4 & 4 \\
\hline
\end{tabular}
\end{center}
\vspace{-3mm}
\caption{The {numbers} of linearly independent zero-mode eigenstates with $Z_4$ eigenvalue $\eta$
  for
  $(\alpha_1,\alpha_2)=\left({1\over 2},{1\over 2}\right)$ 
on $T^2/Z_4$.}
\label{T2Z4Psi_2}
\end{table}

\subsection{$T^2/Z_6$}

{For the case of $T^2/Z_6$}, there is no new future to be declared separately and all the calculations are basically the same with those in the previous $T^2/Z_3$, {even though} it is somewhat more complicated.
Thereby, it suffices to pick up only important points which are different from $T^2/Z_3$ and to write down components of the projective operator in this case.

{From the requirement {(\ref{constraints_on_states_ZN})}}, the two Scherk-Schwarz phases $\alpha_1$ and $\alpha_2$ should comply with the {conditions
\al{
\alpha_1 = \alpha_2 \quad \text{and} \quad
{\alpha_2 = \alpha_2 - \alpha_1 {-} \frac{M}{2}} \quad (\text{mod }1),
}
and} only the following varieties are realizable,
\al{
\alpha &:= \alpha_1 = \alpha_2 \\
\alpha&=
	\begin{cases}
	\displaystyle 0 & \text{for } M : \text{even}, \\[2mm]
	\displaystyle \frac{1}{2} & \text{for } M : \text{odd}. 
	\end{cases}
}

To evaluate the form of the $Z_{6}$-transformed state, we can also consider the following state
\al{
\sum_{j=0}^{M-1} \Usix \Ket{\frac{j}{M}}.
		\label{Z6_relation1}
}
Since the above state satisfies the condition
	\begin{align}
	e^{2\pi i \widehat{\widetilde{Y}}}\sum_{j=0}^{M-1} \Usix \Ket{\frac{j}{M}}= \sum_{j=0}^{M-1} \Usix \Ket{\frac{j}{M}}{,}
	\end{align}
we can reach to the form
	\begin{align}
	\sum_{j=0}^{M-1} \Usix \Ket{\frac{j}{M}}=\A e^{i\theta}\left|0 \right\rangle.
	\end{align}
With casting the shift operator $e^{-i\frac{l}{M}\widehat{\widetilde{P}}}$ on both sides, we obtain
	\begin{align}
	e^{i\frac{\pi}{M}l^2-2\pi i \frac{\alpha}{M}l }\sum^{{M-1}}_{k=0}e^{-2\pi i \frac{l\cdot k}{M}}\widehat{U}_{Z_{6}}\left| \frac{k}{M}\right\rangle=\A e^{i\theta}\left|\frac{l}{M}\right\rangle,
	\end{align}
and subsequently, we can derive the following simple form from the above with summing over $l$ from 0 to $M-1${,}
	\begin{align}
	\widehat{U}_{Z_{6}}\left|0 \right\rangle=\frac{\A}{M} e^{i\theta}\sum_{k=0}^{M-1}e^{-i\frac{\pi}{M}k^2 +2\pi i \frac{\alpha}{M}k}\left|\frac{k}{M}\right\rangle.\label{U6Ket0}
	\end{align}
The $Z_{6}$-consistency $(\widehat{U}_{Z_{6}})^6 =\widehat{1}$ will lead to the following values for $\A$ and $\theta$,
with the help of {Eqs.~(\ref{Trotation_Z6}), (\ref{constraints_on_states_ZN}), (\ref{shift_operaton_on_state}), (\ref{root_formula}), (\ref{magic_formula}), (\ref{U6Ket0})}{,}
\al{
\A = \sqrt{M},\quad
\theta = {\frac{\pi}{12} + \frac{\pi}{M} \alpha^2}.
}
In the above, we {neglected} the trivial overall phase. 

{Summarizing} the above results, we can {indicate the forms of the $Z_{6}$-transformed states} for constructing the projective operator in $T^2/Z_6$:
\al{
\Usix \Ket{\frac{j}{M}} = \left(\Usix^\dagger\right)^5 \Ket{\frac{j}{M}} &=
	{\frac{1}{\sqrt{M}} e^{i\frac{\pi}{12} + i\frac{\pi}{M}\alpha^2  }
	\sum_{k=0}^{M-1} e^{-i\frac{\pi}{M}k^2 +2\pi i\frac{\alpha}{M}k +2\pi i\frac{j\cdot k}{M} } \Ket{\frac{k}{M}}}, \notag \\
\left(\Usix\right)^2 \Ket{\frac{j}{M}} = \left(\Usix^\dagger\right)^4 \Ket{\frac{j}{M}} &=
	{\frac{1}{\sqrt{M}} e^{-i\frac{\pi}{12} + i\frac{3\pi\alpha^2}{M} +i\frac{\pi}{M}j^2 +2\pi i \frac{\alpha}{M}j}
	\sum_{k=0}^{M-1} e^{i\frac{4 \pi\alpha}{M}k +2\pi i \frac{j\cdot k}{M}} \Ket{\frac{k}{M}}},\notag \\
\left(\Usix\right)^3 \Ket{\frac{j}{M}} = \left(\Usix^\dagger\right)^3 \Ket{\frac{j}{M}} &=
	{e^{-i\frac{4\pi\alpha^2}{M} - i\frac{4\pi\alpha}{M} j}
	\Ket{\frac{-2\alpha-j}{M}}}, \notag \\
\left(\Usix\right)^4 \Ket{\frac{j}{M}} = \left(\Usix^\dagger\right)^2 \Ket{\frac{j}{M}} &=
	{\frac{1}{\sqrt{M}} e^{i\frac{\pi}{12} - i\frac{3\pi\alpha^2}{M} - i\frac{4\pi\alpha}{M}j}
	\sum_{k=0}^{M-1} e^{-i\frac{\pi}{M}k^2 -2\pi i\frac{\alpha}{M}k-2\pi i \frac{j\cdot k}{M}} \Ket{\frac{k}{M}}},\notag \\
\left(\Usix\right)^5 \Ket{\frac{j}{M}} = \Usix^\dagger \Ket{\frac{j}{M}} &=
	{\frac{1}{\sqrt{M}} e^{-i\frac{\pi}{12} -i \frac{\pi}{M}\alpha^2{+}i\frac{\pi}{M}j^2 -2\pi i \frac{\alpha}{M}j}
	\sum_{k=0}^{M-1} e^{-2\pi i \frac{j\cdot k}{M}} \Ket{\frac{k}{M}}},
	\label{finalresults_T2Z6}
}
which {are} equivalent to the following expressions:
	\begin{align}
	&\left(\widehat{U}_{Z_{6}}\right)^{x}\left|\frac{j}{M}\right\rangle=\sum_{k=0}^{M-1}D_{jk}^{(\omega^x)}\left|\frac{k}{M}\right\rangle,\hspace{3em}(x=0,1,2,3,4,5),\nonumber\\
	&D_{jk}^{(\omega)}=\frac{1}{\sqrt{M}} e^{i\frac{\pi}{12} + i\frac{\pi}{M}\alpha^2  }e^{-i\frac{\pi}{M}k^2 +2\pi i\frac{\alpha}{M}k +2\pi i\frac{j\cdot k}{M} },\nonumber\\
	&D_{jk}^{(\omega^2)}=\frac{1}{\sqrt{M}} e^{-i\frac{\pi}{12} + i\frac{3\pi\alpha^2}{M} +i\frac{\pi}{M}j^2 +2\pi i \frac{\alpha}{M}j}e^{i\frac{4 \pi\alpha}{M}k +2\pi i \frac{j\cdot k}{M}},\nonumber\\
	&D_{jk}^{(\omega^3)}=e^{-i\frac{4\pi\alpha^2}{M} - i\frac{4\pi\alpha}{M} j}\delta_{{-2\alpha-j,k}},\nonumber\\
	&D_{jk}^{(\omega^4)}=\frac{1}{\sqrt{M}} e^{i\frac{\pi}{12} - i\frac{3\pi\alpha^2}{M} - i\frac{4\pi\alpha}{M}j}e^{-i\frac{\pi}{M}k^2 -2\pi i\frac{\alpha}{M}k-2\pi i \frac{j\cdot k}{M}},\nonumber\\
	&D_{jk}^{(\omega^5)}=\frac{1}{\sqrt{M}} e^{-i\frac{\pi}{12} -i \frac{\pi}{M}\alpha^2{+}i\frac{\pi}{M}j^2 -2\pi i \frac{\alpha}{M}j}e^{-2\pi i \frac{j\cdot k}{M}}.\label{finalresults_T2Z6part2}
	\end{align}

Finally, we construct physical states on $T^2/Z_{6}$ at last. By using {Eq.~(\ref{OperatorPhysicalStates})}, the physical states are represented by the follows:
	\begin{align}
	\left|\frac{j}{M}\right\rangle_{T^2/Z_{6},\eta}&=\frac{1}{6}\sum_{x=0}^{5}\bar{\eta}^{x}\sum_{k=0}^{M-1}D^{(\omega^x)}_{jk}\left|\frac{k}{M}\right\rangle\nonumber\\
	&=:\sum^{M-1}_{k=0}M_{jk}^{(Z_{6};\eta)}\left|\frac{k}{M}\right\rangle,\hspace{4em}(\eta=1,\omega,\omega^2,\omega^3,\omega^4,\omega^5),
	\end{align}
where
	\begin{align}
	M_{jk}^{(Z_{6};\eta)}=\frac{1}{6}\sum_{x=0}^{5}\bar{\eta}^{x}D^{(\omega^x)}_{jk}. \label{Z6M}
	\end{align}
The number of independent physical states $\left|\frac{j}{M}\right\rangle_{T^2/Z_{6},\eta}$ is nothing but the rank of the matrix $M_{jk}^{(Z_{6};\eta)}${,}
	\begin{align}
	\text{The number of physical states }\left|n,\frac{j}{M}\right\rangle_{T^2/Z_{6},\eta}= \text{Rank}\left[M_{jk}^{(Z_{6};\eta)}\right].
	\end{align}
After investigating the rank of the matrix $M_{jk}^{(Z_{6};\eta)}$, we obtain {the results shown in Table~\ref{T2Z6Psi_1} and \ref{T2Z6Psi_2}.} We again emphasize that the following results are consistent with the previous wavefunction analysis {\cite{Abe:2013bca}}.
\begin{table}[htbp]
\vspace{6mm}
\begin{center}
\begin{tabular}{|c|c|ccccccccccccc|}
\hline
\multicolumn{2}{|c|}{$|M|$}& 2 & 4 & 6 & 8 & 10 & 12 & 14 & 16 & 18 & 20 & 22 & 24 & 26 \\
\hline
\multirow{6}{*}{$\eta$}&1 & 1 & 1 & 2 & 2 & 2 & 3 & 3 & 3 & 4 & 4 & 4 & 5 & 5 \\
&$\omega$ & 0 & 1 & 1 & 1 & 2 & 2 & 2 & 3 & 3 & 3 & 4 & 4 & 4 \\
&$\omega^2$ & 1 & 1 & 1 & 2 & 2 & 2 & 3 & 3 & 3 & 4 & 4 & 4 & 5 \\
&$\omega^3$ & 0 & 0 & 1 & 1 & 1 & 2 & 2 & 2 & 3 & 3 & 3 & 4 & 4 \\
&$\omega^4$ & 0 & 1 & 1 & 1 & 2 & 2 & 2 & 3 & 3 & 3 & 4 & 4 & 4 \\
&$\omega^5$ & 0 & 0 & 0 & 1 & 1 & 1 & 2 & 2 & 2 & 3 & 3 & 3 & 4 \\
\hline
\end{tabular}
\end{center}
\vspace{-3mm}
\caption{The {numbers} of linearly independent zero-mode eigenstates with $Z_6$ eigenvalue $\eta$
  for $M=\mathrm{even}$ and
  $(\alpha_1,\alpha_2)=(0, 0)$ 
on $T^2/Z_6$.}
\label{T2Z6Psi_1}
\end{table}
\begin{table}[htbp]
\vspace{6mm}
\begin{center}
\begin{tabular}{|c|c|ccccccccccccc|}
\hline
\multicolumn{2}{|c|}{$|M|$}& 1 & 3 & 5 & 7 & 9 & 11 & 13 & 15 & 17 & 19 & 21 & 23 & 25 \\
\hline
\multirow{6}{*}{$\eta$}&1 & 0 & 1 & 1 & 1 & 2 & 2 & 2 & 3 & 3 & 3 & 4 & 4 & 4 \\
&$\omega$ & 1 & 1 & 1 & 2 & 2 & 2 & 3 & 3 & 3 & 4 & 4 & 4 & 5 \\
&$\omega^2$ & 0 & 0 & 1 & 1 & 1 & 2 & 2 & 2 & 3 & 3 & 3 & 4 & 4 \\
&$\omega^3$ & 0 & 1 & 1 & 1 & 2 & 2 & 2 & 3 & 3 & 3 & 4 & 4 & 4 \\
&$\omega^4$ & 0 & 0 & 0 & 1 & 1 & 1 & 2 & 2 & 2 & 3 & 3 & 3 & 4 \\
&$\omega^5$ & 0 & 0 & 1 & 1 & 1 & 2 & 2 & 2 & 3 & 3 & 3 & 4 & 4 \\
\hline
\end{tabular}
\end{center}
\vspace{-3mm}
\caption{The {numbers} of linearly independent zero-mode eigenstates with $Z_6$ eigenvalue $\eta$
  for $M=\mathrm{odd}$ and
  $(\alpha_1,\alpha_2)=\left({1\over 2},{1\over 2}\right)$ 
on $T^2/Z_6$.}
\label{T2Z6Psi_2}
\end{table}

\section{Correspondence between the operator analysis and the wavefunction analysis
\label{section:analysis}}

{Based on the analytical results that we have obtained in the previous section, {we {will derive} the correspondence between the wavefunction analysis in Section~\ref{section:wavefunction-analysys} and the operator analysis in Section~\ref{section:T2/ZN}.
First, we introduce the description of the magnetized twisted orbifold $T^2/Z_N$ with wavefunctions and rewrite it into the words of the operator analysis. After that, we check the consistency between the operator analysis and the wavefunction analysis.}
}

{First}, we start to discuss {the relation when $M>0$
\al{
f_{T^2,{\Psi_+},0}^{(j+\alpha_1,\alpha_\tau)} (z) = {\Braket{z|0, \frac{j}{M}}^{(\alpha_1,\alpha_\tau)}}{.} \label{wavefunctionconstruction}
}
Although} we explicitly write down the energy eigenvalue and the Scherk-Schwarz phases {in Eq.~(\ref{wavefunctionconstruction}) we hereafter omit} the information on the $T^2$ state $\Ket{0, j/M}^{(\alpha_1,\alpha_\tau)}$ as $\Ket{j/M}$ for simplicity in description.

{The} rotated state $\Ket{\omega z}$ on $T^2/Z_N$ can be expressed with the operator $\widehat{U}_{Z_N}$
\al{
\Ket{\omega z} = {\widehat{U}_{Z_N}^{\dagger}} \Ket{z}.
}
Now, when we consider the following product
\al{
C^{(\omega^{{x}})}_{jk} = 
\int_{T^2} dz d\bar{z}
f_{T^2,{\Psi_+},0}^{({j}+\alpha_1,\alpha_{\tau})} (\omega^{{x}} z)
\left(f_{T^2,{\Psi_+},0}^{({k}+\alpha_1,\alpha_{\tau})} (z) \right)^\ast,\quad
{(x=0,1,\cdots,N-1),}
\label{definition_of_C}
}
it can be represented in the operator formalism as
\al{
C^{(\omega^{{x}})}_{jk} = 
\int_{T^2} dz d\bar{z} \Braket{\frac{{k}}{M} | z} \Braket{z|\left( {\widehat{U}_{Z_N}} \right)^{{x}}| \frac{{j}}{M}}
=
\Braket{\frac{{k}}{M}|\left( {\widehat{U}_{Z_N}} \right)^{{x}} | \frac{{j}}{M}}, 
}
where we {used} the completeness relation $\int_{T^2} dz d\bar{z} \Ket{z}\Bra{z} = 1$.
As we calculated in the previous section, the state $\left( {\widehat{U}_{Z_N}} \right)^{{x}} \Ket{{j/M}}$ can be represented as a linear combination of the states on {$T^2$
\al{
\left( {\widehat{U}_{Z_N}} \right)^{{x}} {\Ket{\frac{{j}}{M}}} = \sum_{m=0}^{{M-1}} D_{{j}m}^{(\omega^{{x}})} \Ket{\frac{m}{M}}.
\label{coefficient_relation}
}
Then, from the {orthonormality} relation $\langle \frac{k}{M} \vert \frac{m}{M}\rangle = \delta_{k,m} $ we can conclude that}
\al{
C^{(\omega^{{x}})}_{jk} = D_{{jk}}^{(\omega^{{x}})},
\label{corresponding_relation_1}
}
where the concrete forms of $D_{{jk}}^{(\omega^{{x}})}$ are found in {Eqs.~(\ref{finalresults_T2Z3part2}), (\ref{finalresults_T2Z4part2}) and (\ref{finalresults_T2Z6part2})} in the {$T^2/Z_3$, $T^2/Z_4$ and $T^2/Z_6$ cases, respectively.}
Since we already obtained all the analytical forms of $D_{{jk}}^{(\omega^{{x}})}$ in the previous section, {we are now} able to express the following relation analytically by use of {$D_{jk}^{(\omega^{x})}$
\al{
f_{T^2,{\Psi_+},0}^{(j+\alpha_1,\alpha_{\tau})} (\omega^{{x}} z) =\sum_{k=0}^{{M-1}}D_{{jk}}^{(\omega^{{x}})}~f_{T^2,{\Psi_+},0}^{({k}+\alpha_1,\alpha_{\tau})} (z). \label{wavefunctionexpansion}
}
We} would like to note that {the} relation in {Eq.~(\ref{wavefunctionexpansion})} itself is interesting because {it brings us lots of nontrivial formulas on the theta functions}.
{By using the numerical calculation for $C^{(\omega^{{x}})}_{jk}$ in Eq.~(\ref{definition_of_C}), we can easily check the validity of the relation Eq.~(\ref{corresponding_relation_1}).}

Next, we try to see the correspondence in the {orbifold} cases directly.
The wavefunctions on $T^2/Z_N$ can be shown in operator formalism by use of the projection operator {$\widehat{\mathcal{P}}_{T^2/Z_N, \eta}$ for the $Z_N$ eigenstate with eigenvalue $\eta$ in Eq.~(\ref{defining_projection_operator})
\al{
f_{T^2/Z_N,{\Psi_+},0}^{(j+\alpha_1,\alpha_\tau)} (z)_{\eta} = \Braket{z| \widehat{\mathcal{P}}_{T^2/Z_N, \eta} |\frac{j}{M}},
}
and the kinetic matrix in Eq.~(\ref{kinetic_matrix}) is rephrased as follows:
\al{
\mathcal{K}_{jk}^{(Z_{N};\eta)} &= \int_{T^2}dzd\bar{z}~
\Braket{\frac{j}{M}| \widehat{\mathcal{P}}_{T^2/Z_N, \eta} |z}
\Braket{z| \widehat{\mathcal{P}}_{T^2/Z_N, \eta} |\frac{k}{M}} \notag \\
&= \Braket{\frac{j}{M}| \widehat{\mathcal{P}}_{T^2/Z_N, \eta} |\frac{k}{M}} \notag \\
&= \Braket{\frac{j}{M}|\frac{1}{N} \sum_{x=0}^{N-1} \bar{\eta}^{x} \left( \widehat{U}_{Z_N} \right)^x |\frac{k}{M}} \notag \\
&=\left\langle\frac{j}{M}\right|\frac{1}{N}\sum^{N-1}_{x=0}\bar{\eta}^{x} \sum^{{M-1}}_{s=0}D^{(\omega^x)}_{ks} \left|\frac{{s}}{M}\right\rangle\notag\\
&={M_{kj}^{(Z_{N}; \eta)}}, \label{KvsM}
}
where} we {used} the completeness relation on $z$ and the {property} of the projection operator, \\
{$\left( \widehat{\mathcal{P}}_{T^2/Z_N, \eta} \right)^2 = \widehat{\mathcal{P}}_{T^2/Z_N, \eta}$} with Eq.~(\ref{coefficient_relation}).
As shown in Eq.~(\ref{KvsM}),
all the materials of {$\mathcal{K}_{jk}^{(Z_{N};\eta)}$} are analytically evaluated and then {we can now} check the consistency {between the results of both the analyses} by calculating the rank of {$\mathcal{K}_{jk}^{(Z_{N};\eta)}$ and form of the unitary matrix diagonalizing $\mathcal{K}_{jk}^{(Z_{N};\eta)}$}.
We {already} re-evaluated the numbers of independent zero-mode eigenstates in all the cases of $T^2/Z_2$, $T^2/Z_3$, $T^2/Z_4$ and $T^2/Z_6$ {in Section~\ref{section:T2/ZN}} with every possible combination of the Scherk-Schwarz {phases. We then} confirmed the agreement {between the results of both the analyses} with exhausting the possibilities of three generations.\footnote{
{Situations are the same with the right-handed KK modes (in the case of $M>0$) since all the discussions are valid irrespective of the KK number.
We can consider the left-handed KK modes (without corresponding zero mode in the case of $M>0$) with the help of the Dirac equation for KK fermionic states.}
}
All the results are totally consistent with the previous ones in Ref.~\cite{Abe:2013bca} and they are summarized {in Tables~\ref{T2Z3Psi_1e} to \ref{T2Z6Psi_2},} where we {skipped} the $T^2/Z_2$ because in this case, exact formulas are available in Ref.~\cite{Abe:2013bca}.


Finally, we show some specific examples of the correspondence.
The first one is the $T^2/Z_6$ case with $M=2$ and $(\alpha_1,\alpha_2) = (0,0)$.
In the $Z_6$ orbifolding, we should consider linear combinations of six terms and calculations requires lots of efforts.
However, {when we utilize the operator formalism on $T^2/Z_6$}, situations get to be very clear.
We can explicitly evaluate the forms of the inner products {$\Bra{\frac{j}{M}} \widehat{\mathcal{P}}_{T^2/Z_6, \eta} \Ket{\frac{k}{M}} \,\left(= \mathcal{K}_{jk}^{(Z_{6};\eta)}={M_{kj}^{(Z_{6}; \eta)}}\right)$} for all the cases of ${x}$ by use of the results {in Eq.~(\ref{finalresults_T2Z6part2}) and Eq.~(\ref{Z6M}).}
These concrete forms {are
\al{
\mathcal{K}^{(Z_{6};\omega^0)}_{{\text{op}}} &=
\left(
\begin{array}{cc}
 \frac{1}{6} \left(2+\sqrt{2} e^{-\frac{i \pi }{12}}+\sqrt{2} e^{\frac{i \pi }{12}}\right) & \frac{1}{6} \left(i \sqrt{2} e^{-\frac{i
   \pi }{12}}+\sqrt{2} e^{\frac{i \pi }{12}}\right) \\
 \frac{1}{6} \left(\sqrt{2} e^{-\frac{i \pi }{12}}-i \sqrt{2} e^{\frac{i \pi }{12}}\right) & \frac{1}{6} \left(2-i \sqrt{2}
   e^{-\frac{i \pi }{12}}+i \sqrt{2} e^{\frac{i \pi }{12}}\right) \\
\end{array}
\right)   \notag \\
&\approx
\left(
\begin{array}{cc}
 {0.789} & {0.289\, +0.289 i} \\
 {0.289\, -0.289 i} & {0.211} \\
\end{array}
\right), \\[0.2cm]
\mathcal{K}^{(Z_{6};\omega^1)}_{{\text{op}}} &=
\left(
\begin{array}{cc}
 0 & 0 \\
 0 & 0 \\
\end{array}
\right), \\[0.2cm]
\mathcal{K}^{(Z_{6};\omega^2)}_{{\text{op}}} &=
\left(
\begin{array}{cc}
 \frac{1}{6} \left(2+\sqrt{2} e^{-\frac{7 i \pi }{12}}+\sqrt{2} e^{\frac{7 i \pi }{12}}\right) & \frac{1}{6} \left(\sqrt{2}
   e^{-\frac{7 i \pi }{12}}+i \sqrt{2} e^{\frac{7 i \pi }{12}}\right) \\
 \frac{1}{6} \left(-i \sqrt{2} e^{-\frac{7 i \pi }{12}}+\sqrt{2} e^{\frac{7 i \pi }{12}}\right) & \frac{1}{6} \left(2+i \sqrt{2}
   e^{-\frac{7 i \pi }{12}}-i \sqrt{2} e^{\frac{7 i \pi }{12}}\right) \\
\end{array}
\right)   \notag \\
&\approx
\left(
\begin{array}{cc}
{0.211} & {-0.289-0.289 i} \\
{-0.289+0.289 i} & {0.789} \\
\end{array}
\right), \\[0.2cm]
\mathcal{K}^{(Z_{6};\omega^3)}_{{\text{op}}} &=
\left(
\begin{array}{cc}
 0 & 0 \\
 0 & 0 \\
\end{array}
\right), \\[0.2cm]
\mathcal{K}^{(Z_{6};\omega^4)}_{{\text{op}}} &=
\left(
\begin{array}{cc}
 0 & 0 \\
 0 & 0 \\
\end{array}
\right), \\[0.2cm]
\mathcal{K}^{(Z_{6};\omega^5)}_{{\text{op}}} &=
\left(
\begin{array}{cc}
 0 & 0 \\
 0 & 0 \\
\end{array}
\right),
}
where} the subscript ``op" indicates that these matrices are evaluated in operator formalism explicitly.}
The above result means that there is no physical degree of freedom in the cases of {$\eta=\omega^{1},\omega^{3},\omega^{4},\omega^{5}$}.
For {$\eta=\omega^{0},\omega^2$, the unitary matrix $U$ diagonalizing the ``kinetic matrix" $\mathcal{K}^{(Z_{6};\eta)}$} can be easily calculated,
\al{
U |_{{x}=0} &=
{\left(
\begin{array}{cc}
 \sqrt[4]{\frac{1}{6} \left(-2-\sqrt{3}\right)} & -\sqrt[4]{\frac{1}{6} \left(-2+\sqrt{3}\right)} \\
 \frac{1}{\sqrt{3+\sqrt{3}}} & \sqrt{\frac{1}{6} \left(3+\sqrt{3}\right)} \\
\end{array}
\right)}, \\
U |_{{x}=2} &=
{\left(
\begin{array}{cc}
 -\sqrt[4]{\frac{1}{6} \left(-2+\sqrt{3}\right)} & \sqrt[4]{\frac{1}{6} \left(-2-\sqrt{3}\right)} \\
 \sqrt{\frac{1}{6} \left(3+\sqrt{3}\right)} & \frac{1}{\sqrt{3+\sqrt{3}}} \\
\end{array}
\right)}.
}
After the manipulation {$U^\dagger \mathcal{K}^{(Z_{6};\eta)} U$}, the kinetic terms are suitably diagonalized as follows:
\al{
U^\dagger \mathcal{K}^{(Z_{6};\omega^0)}_{{\text{op}}} U =
\left(
\begin{array}{cc}
 1 & 0 \\
 0 & 0 \\
\end{array}
\right), \quad
U^\dagger \mathcal{K}^{(Z_{6};\omega^2)}_{{\text{op}}} U =
\left(
\begin{array}{cc}
 1 & 0 \\
 0 & 0 \\
\end{array}
\right).
}
The two expressions clearly {indicate} that one mode is physical after the orbifolding in ${x}=0,2$.

{Here,} we comment on consistency with the numerical calculation with {the explicit forms of the theta functions}.
{By} calculating Eq.~(\ref{kinetic_matrix}) numerically with the theta functions in Eq.~(\ref{fermion_mode_function}), we obtain the following corresponding results{,
\al{
\mathcal{K}^{(Z_{6};\omega^0)}_{{\text{wf}}} &=
{\left(
\begin{array}{cc}
 {0.789}& {0.289\, +0.289 i} \\
 {0.289\, -0.289 i} & {0.211} \\
\end{array}
\right)}, \\
\mathcal{K}^{(Z_{6};\omega^1)}_{{\text{wf}}} &=
\left(
\begin{array}{cc}
 0 & 0 \\
 0 & 0 \\
\end{array}
\right), \\
\mathcal{K}^{(Z_{6};\omega^2)}_{{\text{wf}}} &=
{\left(
\begin{array}{cc}
 {0.211} & {-0.289-0.289 i} \\
 {-0.289+0.289 i} & {0.789} \\
\end{array}
\right)}, \\
\mathcal{K}^{(Z_{6};\omega^3)}_{{\text{wf}}} &=
\left(
\begin{array}{cc}
 0 & 0 \\
 0 & 0 \\
\end{array}
\right), \\
\mathcal{K}^{(Z_{6};\omega^4)}_{{\text{wf}}} &=
\left(
\begin{array}{cc}
 0 & 0 \\
 0 & 0 \\
\end{array}
\right), \\
\mathcal{K}^{(Z_{6};\omega^5)}_{{\text{wf}}} &=
\left(
\begin{array}{cc}
 0 & 0 \\
 0 & 0 \\
\end{array}
\right),
}
where} the subscript ``wf" is the counterpart of the subscript ``op" in the numerical calculation with wavefunctions {following the approach in Ref.~\cite{Abe:2013bca}}.
We can conclude that they agree with each other {within the error of the numerical computation} as
\al{
{
\mathcal{K}^{(Z_{6};\eta)}_{{\text{op}}} =
\mathcal{K}^{(Z_{6};\eta)}_{{\text{wf}}}
\quad
\text{for } \eta=\omega^{0},\omega^{1},\cdots,\omega^{5},}
}
{and hence, the relation Eq.~(\ref{KvsM}) is valid.}

As the second example, we focus on the $T^2/Z_4$ case with $M=2$ and $(\alpha_1,\alpha_2) = (0,0)$ since the analytical result was already discussed {with the exact forms of the mode functions with {theta functions} as in Eq.~(\ref{fermion_mode_function}) and some related mathematical relations, and their explicit forms are} available in Appendix~C of Ref.~\cite{Abe:2013bca}.
The explicit shapes of the kinetic matrix in {$\eta=\omega^{0},\omega^{1},\omega^{2},\omega^{3}$} {are
\al{
\mathcal{K}^{(Z_{4};\omega^0)}_{{\text{op}}} &=
\left(
\begin{array}{cc}
 \frac{1}{4} \left(2+\sqrt{2}\right) & \frac{1}{2 \sqrt{2}} \\
 \frac{1}{2 \sqrt{2}} & \frac{1}{4} \left(2-\sqrt{2}\right) \\
\end{array}
\right), \\
\mathcal{K}^{(Z_{4};\omega^1)}_{{\text{op}}} &=
\left(
\begin{array}{cc}
 0 & 0 \\
 0 & 0 \\
\end{array}
\right), \\
\mathcal{K}^{(Z_{4};\omega^2)}_{{\text{op}}} &=
\left(
\begin{array}{cc}
 \frac{1}{4} \left(2-\sqrt{2}\right) & -\frac{1}{2 \sqrt{2}} \\
 -\frac{1}{2 \sqrt{2}} & \frac{1}{4} \left(2+\sqrt{2}\right) \\
\end{array}
\right), \\
\mathcal{K}^{(Z_{4};\omega^3)}_{{\text{op}}} &=
\left(
\begin{array}{cc}
 0 & 0 \\
 0 & 0 \\
\end{array}
\right).
}
Apparently}, physical modes survive only in the cases {$\eta=\omega^{0},\omega^{2}$}.
Each unitary matrix for diagonalizing the kinetic matrix takes the following form:
\al{
U|_{{x}=0} &=
\left(
\begin{array}{cc}
 \frac{\sqrt{2+\sqrt{2}}}{2} & -\frac{1}{2} \sqrt{2-\sqrt{2}} \\
 \frac{1}{\sqrt{2 \left(2+\sqrt{2}\right)}} & \frac{1}{\sqrt{4-2 \sqrt{2}}} \\
\end{array}
\right), \\
U|_{{x}=2} &=
\left(
\begin{array}{cc}
 -\frac{1}{2} \sqrt{2-\sqrt{2}} & \frac{\sqrt{2+\sqrt{2}}}{2} \\
 \frac{1}{\sqrt{4-2 \sqrt{2}}} & \frac{1}{\sqrt{2 \left(2+\sqrt{2}\right)}} \\
\end{array}
\right),
}
and the diagonalized kinetic terms are calculated {as
\al{
U^\dagger \mathcal{K}^{(Z_{4};\omega^0)}_{{\text{op}}} U =
\left(
\begin{array}{cc}
 1 & 0 \\
 0 & 0 \\
\end{array}
\right), \quad
U^\dagger \mathcal{K}^{(Z_{4};\omega^2)}_{{\text{op}}} U =
\left(
\begin{array}{cc}
 1 & 0 \\
 0 & 0 \\
\end{array}
\right).
}
After} evaluating the ratios of matrix elements of $U$ as
\al{
\frac{U_{21}|_{{x}=0}}{U_{11}|_{{x}=0}} = \sqrt{2}-1, \quad
\frac{U_{21}|_{{x}=2}}{U_{11}|_{{{x}=2}}} = -\sqrt{2}-1,
}
these values should correspond to the ratios of coefficients in the construction with mode functions on $T^2$ discussed in Ref.~\cite{Abe:2013bca}.
We can easily check that this statement is correct.

\section{Summary and Discussions
\label{section:summary}}

We have discussed an effective way for analyzing the system on the magnetized twisted orbifolds in operator formalism, especially in the following complicated cases $T^2/Z_3, T^2/Z_4, T^2/Z_6$.
With the help of {mathematical formulas}, we have obtained the exact and analytical results which can be applicable for any larger values of the quantized magnetic flux $M$.
The (non-diagonalized) kinetic terms are immediately generated via the formalism and the number of the surviving physical states {are} straightforwardly calculable in a rigorous manner by simply following usual procedures in linear algebra.
We {have} checked and re-derived all the results in Ref.~\cite{Abe:2013bca} based on huge numerical computations with ease analytically.

Based on the achievement in this paper, we can consider a few next directions.
One is to construct actual (semi-)realistic models based on the magnetized twisted orbifolds of $T^2/Z_2$, $T^2/Z_3$, $T^2/Z_4$ and $T^2/Z_6$.
Even in the simplest $T^2/Z_2$ case, possibilities with nontrivial Scherk-Schwarz phases are not yet touched.
Complex geometries would help us to generate the complicated nature of fermion flavor structure in the SM, and also to eliminate unwanted exotic states in the zero-mode sector.
Here, we would like to emphasize that all the technical obstacles in analysis were removed by the formalism which we {have discussed}.
{It is also interesting to study non-Abelian flavor symmetries 
appearing in $T^2/Z_N$ orbifolds with magnetic fluxes~\cite{Abe:2009vi,Abe:2009uz,Abe:2010ii,BerasaluceGonzalez:2012vb,Honecker:2013hda,Marchesano:2013ega}.
(See also~\cite{Hamada:2014hpa}.)}

Another option is to analyze other geometries, {{\it e.g.}}, magnetized twisted orbifolds based on higher-dimensional torus in operator formalism {\cite{Sakamoto:2003rh}}. 
Our strategy which has been used in this paper is expected to be valid in such much more complicated cases.
``Dualities" between analyses with wavefunctions and {operator formalism} are quite interesting since we can excavate mathematical relations like in Eqs.~(\ref{definition_of_C}) and (\ref{corresponding_relation_1}).
{Such a study} can be a fascinating theme in mathematical physics.


\section*{Acknowledgments}

We would like to thank Hiroto So for helping us to prove some essential mathematical relations, whose detail is shown in Appendix~\ref{appendix:So_formula}, used in analysis.
We are also grateful for valuable discussions with Masazumi~Honda, Yoshinori~Honma, Shogo~Tanimura {and Satoshi Yamaguchi. We also thank Yoshiyuki~Tatsuta for suggesting a typo in the first version of this manuscript.}
This work was supported in part by scientific grants from the Ministry of Education, Culture,
Sports, Science and Technology under Grants No.~25$\cdot$3825 (Y.F.), No.~25400252 (T.K.) and No.~20540274 (M.S.).
K.N. is partially supported by funding available from the
Department of Atomic Energy, Government of India for the Regional
Centre for Accelerator-based Particle Physics (RECAPP), Harish-Chandra
Research Institute.


\appendix
\section*{Appendix}
\section{Large gauge transformation in operator formalism
\label{appendix:gauge}}

Based on the knowledge obtained in section~\ref{section:T2}, we reconsider the quantum mechanical system on $T^2/Z_N$ as
\al{
{\widehat{H} \Ket{\psi} = E \Ket{\psi},\quad
e^{i \widehat{T}_a} \Ket{\psi} = e^{2 \pi i \alpha_a} \Ket{\psi}\quad
(a=1,2),}
}
where the Hamiltonian {$\widehat{H}$} and the operator {$\widehat{T}_a$} describing the constraints on the state on $T^2/Z_N$, $\Ket{\psi}$, are given as
\al{
{\widehat{H} = \left( \widehat{\mathbf{p}} {+} \frac{q}{2} \Omega \left( \widehat{\mathbf{y}} + \mathbf{a} \right) \right)^2,\quad
\widehat{T}_a = \mathbf{u}_a^{\mathrm{T}} \left( \widehat{\mathbf{p}} {+} \frac{q}{2} \Omega^{\mathrm{T}} (\widehat{\mathbf{y}}+\mathbf{a} )\right),}
}
where the gauge-fixing is already done as in Eq.~(\ref{gauge_fixing}).
{$\alpha_1$ and $\alpha_2$} represent the Scherk-Schwarz phases in the original coordinate.

When we examine the gauge transformation,
\al{
\psi(\mathbf{y}) = e^{{-}i \frac{q}{2} {\mathbf{y}^{\mathrm{T}}}(\Omega  \mathbf{a})} \psi'(\mathbf{y}),\quad \Braket{\mathbf{y}|\psi} = \psi(\mathbf{y}),
}
the {Schr\"odingier equation (\ref{original_Hamiltonian}) and the pseudo-periodic boundary condition (\ref{constraint_on_torus_wavefunction}) are modified as
	\al{
	\left(-i\nabla+\frac{q}{2}\Omega\mathbf{y}\right)^2\psi'(\mathbf{y})&=E\psi'(\mathbf{y}),
	}
	\al{
	\psi'(\mathbf{y} + \mathbf{u}_a) &=e^{{-}i \frac{q}{2} (\mathbf{y}+\mathbf{0})^{\mathrm{T}} \Omega \mathbf{u}_a + 2\pi i \alpha_a-iq\mathbf{a}^{\mathrm{T}}\Omega \mathbf{u}_{a}} \psi'(\mathbf{y})\nonumber\\
	&{=:} \, e^{{-}i \frac{q}{2} (\mathbf{y}+\mathbf{0})^{\mathrm{T}} \Omega \mathbf{u}_a + 2\pi i \alpha'_a} \psi'(\mathbf{y})\quad \text{for } a=1,2,
	}
and the forms of the operators {$\widehat{H}$ and $\widehat{T}_a$ }get morphed as follows:
\al{
{\widehat{H} \rightarrow \widehat{H'} =
	\left( \widehat{\mathbf{p}} + \frac{q}{2} \Omega \widehat{\mathbf{y}} \right)^2,\quad
\widehat{T}_a \rightarrow \widehat{T'_a} =
	\mathbf{u}_a^{\mathrm{T}} \left( \widehat{\mathbf{p}} + \frac{q}{2} \Omega^{\mathrm{T}} \widehat{\mathbf{y}} \right) ,}}
where the Wilson line phases are gauged away from the Hamiltonian {$\widehat{H}'$} and the translational operators {$\widehat{T}'_{a}$}.}
Besides, due to the modification in the operators {$\widehat{T'_a}$}, the values of the Scherk-Schwarz phases in the gauge-transformed system {$\alpha'_1$ and $\alpha'_2$} turn out to be
\al{
{2 \pi  \alpha'_a = 2 \pi  \alpha_a {-}  q \,  \mathbf{a}^{\mathrm{T}}\Omega\mathbf{u}_a,}\quad\text{ mod $2\pi$.}
	\label{large_gauge_transformation_on_phase}
}
{The above expression is equivalent to the following expression:
	\al{
	&\left\{ 
	\begin{array}{l}
	2\pi \alpha'_{1}=2\pi\alpha_{1}+2\pi Ma_{2},\\
	2\pi \alpha'_{2}=2\pi\alpha_{2}-2\pi Ma_{1},
	\end{array}
	\right. \hspace{7em} \text{for $T^2/Z_{2},$}\nonumber\\[0.3cm]
	&\left\{
	\begin{array}{l}
	2\pi \alpha'_{1}=2\pi\alpha_{1}+\frac{2\pi}{\sin(2\pi /N)}Ma_{2},\\
	2\pi \alpha'_{2}=2\pi\alpha_{2}-\frac{2\pi}{\sin(2\pi /N)}\biggl( Ma_{1} \sin(2\pi /N) -Ma_{2}\cos(2\pi /N)\biggr),
	\end{array}
	\right. \nonumber\\
	&\hspace{19em}{\text{for $T^2/Z_{N}$ ($N=3,4,6$)},}\label{large_gauge_transformation_on_phase2}
	}
where we used the choices of $\mathbf{u}_1=(1,0)^\mathrm{T}$, $\mathbf{u}_2=(0,1)^\mathrm{T}$ for $N=2$ and $\mathbf{u}_1=(1,0)^\mathrm{T}$, $\mathbf{u}_2=(\cos(2\pi/N),\sin(2\pi/N))^\mathrm{T}$ for $N=3,4,6$, respectively. We also used 
	\al{
	\Omega=\frac{2\pi M}{q\sin(2\pi/N)}\left(\begin{array}{cc}
									0&1\\
									-1&0
									\end{array}\right),
	}
which is obtained  from Eq.~(\ref{magnetic_flux_quantization}) with $M:=Q_{12}$.

Here, we understand that the Wilson line phases and the Scherk-Schwarz phases are correlated under the large gauge transformation and not independent degrees of freedom.
{Now, we conclude that we can take 
\al{
\mathbf{a}=\mathbf{0},
}
without any loss of generality.

When we take {$\alpha_a = 0$}, we can see the correspondence in a direct way. }
Thus, from Eq.~(\ref{large_gauge_transformation_on_phase2}), {the following relations should hold
\al{
Z_2 \text{ case:}\quad Ma_1 &= -\alpha'_2,&
	Ma_2 &= \alpha'_1,\\
Z_N \text{ case:}\quad Ma_1 &= \alpha'_{1}\cos(2\pi /N)-\alpha'_{2},&
	Ma_2 &= \alpha'_{1}\sin(2\pi /N),\quad (N=3,4,6),
}
or
\al{
Z_2 \text{ case:}\quad Ma_1 &= {-\alpha'_2},&
	Ma_2 &= {\alpha'_1},\\
Z_3 \text{ case:}\quad Ma_1 &= -\frac{3}{2} \alpha' ,&
	Ma_2 &=  \frac{\sqrt{3}}{2} \alpha',\\
Z_4 \text{ case:}\quad Ma_1 &= -\alpha',&
	Ma_2 &= \alpha',\\
Z_6 \text{ case:}\quad Ma_1 &= {-\frac{1}{2}} \alpha',&
	Ma_2 &= {\frac{\sqrt{3}}{2}} \alpha'{.}
}
As} we discussed in section~\ref{section:T2/ZN} in the system with $\mathbf{a} = \mathbf{0}$, the two Scherk-Schwarz phases should take the same value {$\alpha' := \alpha'_1 = \alpha'_2$} on $T^2/Z_3$, $T^2/Z_4$ and $T^2/Z_6$.
After we reflect on the fact that {$\alpha'_1$ and $\alpha'_2$} have the period $1$,
we check that the differences of the allowed values of $M(a_1,a_2)$ correspond to the positions of the fixed points of $T^2/Z_N$, which strongly indicate that the number of the {allowed} Scherk-Schwarz phases is connected to the number of the fixed points:
\al{
Z_2 \text{ case:}\quad M(a_1,a_2) &= (0,0), & & (\alpha'_1=0, \alpha'_2=0), \notag \\
		&= (1/2,0), & & (\alpha'_1=1/2, \alpha'_2=0), \notag \\
		&= (0,1/2), & & (\alpha'_1=0, \alpha'_2=1/2), \notag \\
		&= (1/2,1/2), & & (\alpha'_1=1/2, \alpha'_2=1/2){,}\\[5pt]
Z_3 \text{ case:}\quad M(a_1,a_2) &= (0,0), & & (M:\text{even,}\ \alpha'=0), \notag \\
		&= ( {1}/{2}, \sqrt{3}/{6} ), & & (M:\text{even,}\ \alpha'=2/6), \notag \\
		&= ( 0, \sqrt{3}/{3} ), & & (M:\text{even,}\ \alpha'=4/6), \notag \\
		&= ( {3}/{4}, \sqrt{3}/{12} ), & & (M:\text{odd,}\ \alpha'=1/6), \notag \\
		&= ( {1}/{4}, \sqrt{3}/{4} ), & & (M:\text{odd,}\ \alpha'=3/6), \notag \\
		&= ( -{1}/{4}, 5\sqrt{3}/{12} ), & & (M:\text{odd,}\ \alpha'=5/6){,}\\[5pt]
Z_4 \text{ case:}\quad M(a_1,a_2) &= (0,0), & & (\alpha'=0), \notag \\
		&= ( {1}/{2}, 1/{2} ), & & (\alpha'=1/2){,}\\[5pt]
Z_6 \text{ case:}\quad M(a_1,a_2) &= (0,0), & & (M:\text{even,}\ \alpha'=0), \notag \\
		&= ( {3}/{4}, \sqrt{3}/{4} ), & & (M:\text{odd,}\ \alpha'=1/2).
}
Note that the results are totally consistent with those in Ref.~\cite{Abe:2013bca} with wavefunctions.}

\section{Derivation of formulas
\label{appendix:derivation}}

\subsection{Eq.~(\ref{magic_formula})}

In this part, we derive the formula {(\ref{magic_formula})}
\al{
{I_{\pm}} (t,\beta) := \sum_{s=0}^{M-1} e^{-\pi i \frac{(s+t {\pm} \beta)^2}{M}} = \sqrt{M} e^{-\frac{1}{4} \pi i}
\quad
\text{for } t \in \mathbb{Z},\ \beta = 
\begin{cases} \displaystyle 0 & \text{for } M \text{: even}, \\[2mm]
\displaystyle \frac{1}{2} & \text{for } M \text{: odd}, \end{cases}
}
where {$M$ is a positive integer and} the resultant form is independent of {$t$, $\beta$ and the sign in front of $\beta$}.
First, we show the relation ${I_{\pm}}(t,\beta) = {I_{\pm}}(0,\beta)$, which implies the independence of $t$.
\al{
{I_{\pm}}(t,\beta) &= \sum_{s'=t}^{M-1+t} e^{-\pi i \frac{(s' {\pm} \beta)^2}{M}} \quad
(s' := s+t) \notag \\
&= \left[ \sum_{s'=t}^{M-1} + \sum_{s'=M}^{M-1+t} \right] e^{-\pi i \frac{(s' {\pm} \beta)^2}{M}} \notag \\
&= \sum_{s'=t}^{M-1} e^{-\pi i \frac{(s' {\pm} \beta)^2}{M}} + 
\sum_{s''=0}^{t-1} e^{-i \frac{\pi}{M} \left[ (s' {\pm} \beta)^2 + M (M {\pm} 2\beta+2s'') \right]}
\quad (s'' := s' - M).
}
After noticing that $M {\pm} 2\beta$ is always an even integer, we can justify the manipulation,
\al{
e^{-i \frac{\pi}{M} \left[ M (M {\pm} 2\beta+2s'') \right]} =
e^{-i {\pi} (M {\pm} 2\beta+2s'') } \to 1.
}
Then, the following result is obtained
\al{
{I_{\pm}}(t,\beta) = \left[ \sum_{s'=t}^{M-1} + \sum_{s'=0}^{t-1} \right] 
e^{-\pi i \frac{(s' {\pm} \beta)^2}{M}}
= \sum_{s=0}^{M-1} e^{-\pi i \frac{(s {\pm} \beta)^2}{M}} = {I_{\pm}}(0,\beta).
}
From now on, we can set $t=0$ without loss of generality and examine the $\beta=0$ ($M$ is even) and $\beta=1/2$ ($M$ is odd) separately.

The former case {($\beta=0$)} is evaluated straightforwardly.
Using the periodicity of the exponential functions (when $M$ is even), the following deformation is realized
\al{
{I_{\pm}}(0,0) &= \frac{1}{2} \left[ \sum_{k=0}^{M-1} e^{-\pi i \frac{k^2}{M}} +
\sum_{k=M}^{2M-1} e^{-\pi i \frac{(k-M)^2}{M}} \right] \notag \\
&= \frac{1}{2} \left[ \sum_{k=0}^{M-1} + \sum_{k=M}^{2M-1} \right] e^{-\pi i \frac{k^2}{M}}
\notag \\
&= \frac{1}{2} \sum_{k=0}^{N-1} e^{-2\pi i \frac{k^2}{N}} \quad (N:=2M).
}
After using the mathematical relation, which is proved in the next subsection,
\al{
\sum_{s=0}^{N-1} e^{-2\pi i \frac{s^2}{N}} = \sqrt{\frac{N}{2}} e^{-\frac{1}{4}\pi i}
\left( 1 + e^{\frac{N}{2} \pi i} \right) \quad (N \in \mathbb{N}),
		\label{So_formula}
}
we can reach the final form,
\al{
{I_{\pm}}(0,0) &= \frac{1}{2} \sqrt{M} e^{-\frac{1}{4}\pi i}
\left( 1 + e^{M \pi i} \right) = \sqrt{M} e^{-\frac{1}{4}\pi i},
}
where $M$ is even and then $e^{M \pi i} = 1$.

In the latter case {($\beta=1/2$)} is somewhat complicated.
{By use of the fact that $2M$ is even, the following deformation is possible,
\al{
I_{+}\left(0,\frac{1}{2}\right) &= \frac{1}{2} \sum_{s=0}^{N-1} e^{-2\pi i \frac{(2s + 1)^2}{4N}} \quad (N:=2M) \notag \\
&= \frac{1}{2} \sum_{k=+1, \atop k:\text{odd}}^{2N-1} e^{-2\pi i \frac{k^2}{4N}} \quad
(k:=2s+1) \notag \\
&= \frac{1}{2} \left\{ \left[ \sum_{k=+1}^{2N-1} - \sum_{k=+1, \atop k:\text{even}}^{2N-1} \right] e^{-2\pi i \frac{k^2}{4N}} \right\} \notag \\
&= \frac{1}{2} \left\{ \left[ \sum_{k=0}^{2N-1} - \sum_{k=0, \atop k:\text{even}}^{2N-1} \right] e^{-2\pi i \frac{k^2}{4N}} \right\} \notag \\
&= \frac{1}{2} \left\{ \sum_{k=0}^{2N-1} e^{-2\pi i \frac{k^2}{4N}} - \sum_{l=0}^{N-1} e^{-2\pi i \frac{l^2}{N}} \right\} \quad \left(l := \frac{k}{2}\right),
		\label{A1_relation1} \\
{I_{-}}\left(0,\frac{1}{2}\right) &= \frac{1}{2} \sum_{s=0}^{N-1} e^{-2\pi i \frac{(2s-1)^2}{4N}} \quad (N:=2M) \notag \\
&= \frac{1}{2} \sum_{k=-1, \atop k:\text{odd}}^{2N-3} e^{-2\pi i \frac{k^2}{4N}} \quad
(k:=2s-1) \notag \\
&= \frac{1}{2} \left\{ \left[ \sum_{k=-1}^{2N-1} - \sum_{k=-1, \atop k:\text{even}}^{2N-1} \right] e^{-2\pi i \frac{k^2}{4N}} - e^{-2\pi i \frac{(2N-1)^2}{4N}} \right\} \notag \\
&= \frac{1}{2} \left\{ \left[ \sum_{k=0}^{2N-1} e^{-2\pi i \frac{k^2}{4N}} + e^{-2\pi i \frac{(-1)^2}{4N}} \right] - \sum_{k=0, \atop k:\text{even}}^{2N-2} e^{-2\pi i \frac{k^2}{4N}} - e^{-2\pi i \frac{1}{4N} \left[4N(N-1) +1\right]} \right\} \notag \\
&= \frac{1}{2} \left\{ \sum_{k=0}^{2N-1} e^{-2\pi i \frac{k^2}{4N}} - \sum_{l=0}^{N-1} e^{-2\pi i \frac{l^2}{N}} \right\} \quad \left(l := \frac{k}{2}\right),
		\label{A1_relation2}
}
where the two final forms get to be the same.
Here, {the second term of the last line of Eq.~(\ref{A1_relation1}) or (\ref{A1_relation2})} can be calculated with the help of the formula in Eq.~(\ref{So_formula})}{. The first term of the last line in Eq.~(\ref{A1_relation1}) or (\ref{A1_relation2})} needs additional transformations to be evaluated,
\al{
\sum_{k=0}^{2N-1} e^{-2\pi i \frac{k^2}{4N}} &=
	\frac{1}{2} \left[ \sum_{k=0}^{2N-1} e^{-2\pi i \frac{k^2}{4N}} + \sum_{k=2N}^{4N-1} e^{-2\pi i \frac{(k-2N)^2}{4N}} \right] \notag \\
&= \frac{1}{2} \left[ \sum_{k=0}^{2N-1} e^{-2\pi i \frac{k^2}{4N}} + \sum_{k=2N}^{4N-1} e^{-2\pi i \frac{(k^2 + 4N(N-k))}{4N}} \right] \notag \\
&= \frac{1}{2} \left[ \sum_{k=0}^{2N-1} + \sum_{k=2N}^{4N-1} \right] e^{-2\pi i \frac{k^2}{4N}} \notag \\
&= \frac{1}{2} \sum_{k=0}^{4N-1} e^{-2\pi i \frac{k^2}{4N}},
}
in which the final form is applicable for the formula in Eq.~(\ref{So_formula}).
Now, we can show the final result as
\al{
{I_{\pm}}\left(0,\frac{1}{2}\right) &= \frac{1}{4} \sqrt{4M} e^{-\frac{1}{4}\pi i} \left(1+e^{4M\pi i}\right)
-\frac{1}{2} \sqrt{M} e^{-\frac{1}{4}\pi i} \left(1+e^{M\pi i}\right) \notag \\
&= \sqrt{M} e^{-\frac{1}{4}\pi i},
}
where we use the oddness of $M$ as $e^{4M\pi i} = 1,\ e^{M\pi i} = -1$.

\subsection{Eq.~(\ref{So_formula}) \label{appendix:So_formula}}

\begin{figure}[t]
\centering
\includegraphics[width=0.40\columnwidth]{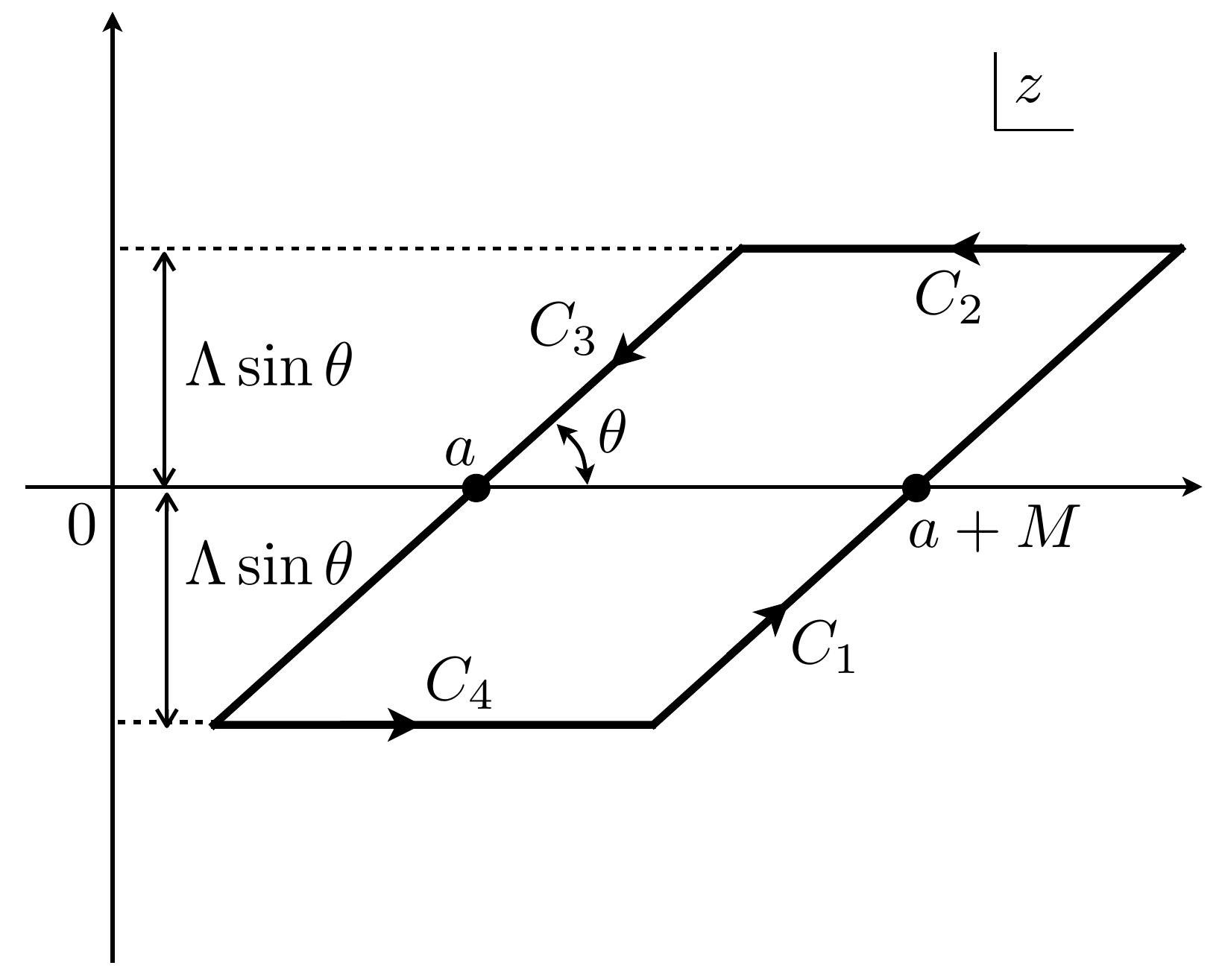}
\caption{
A description for the integral contour consisting of four paths $C_1$, $C_2$, $C_3$ and $C_4$ {with the condition on $\theta$ of $0 < \theta < \pi/4$}.
}
\label{contour_pdf}
\end{figure}

In this part, we prove an essential formula in the previous subsection. 
Firstly, we consider the following function with a positive integer $M$,
\al{
F(z) = \frac{e^{2\pi i z^2/M}}{e^{2\pi i z} -1},
}
where it contains the shift properties,
\al{
F(z+M) &= e^{4\pi i z} F(z), \notag \\
F(z+M) - F(z) &= e^{2\pi i z^2/M + 2\pi i z} + e^{2\pi i z^2/M}.
		\label{A2_shift_property}
}
The integral contour is considered in Fig.~\ref{contour_pdf}, which consists of four paths $C_1$, $C_2$, $C_3$ and $C_4$ {with the condition on $\theta$ of $0 < \theta < \pi/4$}.
Here, we set $a \not\in \mathbb{Z}$ to avoid poles of $F(z)$.

The complex integrals on $C_2$ and $C_4$ can be ignored in the limit $\Lambda \to \infty$
{since $F(z)$ gets to be zero in this limit (when $0 < \theta < \pi/4$),}
and then the remaining integrals are
\al{
I(\Lambda) := \int_{C_1} dz F(z) + \int_{C_3} dz F(z) =
	\int_{-\Lambda}^{\Lambda} dr e^{i\theta} \left[ F(M+a+re^{i\theta}) - F(a+re^{i\theta}) \right].
}
After using the shift properties in Eq.~(\ref{A2_shift_property}), we obtain
\al{
I(\Lambda) = \int_{-\Lambda}^{\Lambda} dr e^{i\theta}
	\left[ e^{2\pi i (a+re^{i\theta})^2/M + 2\pi i (a+re^{i\theta})} +
	e^{2\pi i (a+re^{i\theta})^2/M} \right].
}
When we take a notice of $\sin{2\theta}/M > 0$, changing variable from $r$ to $x:=(re^{i\theta}+a)e^{\pi i/4}$ is available to perform the integration $I(\Lambda)$.
{Also, we can use} generalized Fresnel integrals,
\al{
\int_0^{\infty} dx e^{-(b^2 \cot{\phi})x^2} \cos(b^2 x^2) &=
	\frac{\sqrt{\pi}}{2b} \sqrt{\sin{\phi}} \cos(\phi/2),\\
\int_0^{\infty} dx e^{-(b^2 \cot{\phi})x^2} \sin(b^2 x^2) &=
	\frac{\sqrt{\pi}}{2b} \sqrt{\sin{\phi}} \sin(\phi/2),
}
where $b>0$ and $0<\phi<\pi/2$.

The limiting value
\al{
I(\Lambda \to \infty) = \sqrt{\frac{M\pi}{2\pi}} e^{i\pi/4 -i\pi M/2} + \sqrt{\frac{M\pi}{2\pi}} e^{i\pi/4}
}
is independent of $a$.
Since the poles of $F$ in the contour are located in $z=[a]+1,\cdots,[a]+M$, we use the residue theorem on complex integral,
\al{
\frac{1}{2\pi i} \sum_{k=[a]+1}^{[a]+M} (2\pi i) e^{2\pi i k^2/M} =
	\sqrt{\frac{M\pi}{2\pi}} e^{i\pi/4 -i\pi M/2} + \sqrt{\frac{M\pi}{2\pi}} e^{i\pi/4}.
		\label{A2_relation1}
}
Here, the right-hand side of Eq.~(\ref{A2_relation1}) is independent of $a$ and still,
$e^{2\pi i ([a]+M+1)^2/M} = e^{2\pi i ([a]+1)^2/M}$.
When we set $a$ in the range of $-1<a<0$, the following sum formula can be derived,
\al{
\sum_{k=0}^{M-1} e^{2 \pi i k^2/M} = \sqrt{\frac{M}{2}} e^{i \pi/4 -i\pi M/2} + \sqrt{\frac{M}{2}} e^{i \pi/4} = \sqrt{\frac{M}{2}} e^{i \pi/4} \left( 1 + e^{-\pi i M/2} \right),
}
which is just (complex-conjugated) what we would like to show.




\bibliographystyle{utphys}
\bibliography{operator_magneticflux}


\end{document}